\documentclass[article]{arxiv_manifolds}

\leadauthor{Tostado} 

\title{Neural population dynamics in songbird RA and HVC during learned motor-vocal behavior}

\author[1,2,3$\dagger$]{Pablo Tostado-Marcos}
\author[3$\dagger$]{Ezequiel M. Arneodo}
\author[5]{Lauren Ostrowski}
\author[2,3]{Daril E. Brown II}
\author[2]{Xavier A. Perez}
\author[2]{Adam Kadwory}
\author[5]{Lauren L. Stanwicks}
\author[2]{Abdullah Alothman}
\author[3,4,5$\blacklozenge$]{Timothy Q. Gentner}
\author[2,5$\blacklozenge$]{Vikash Gilja}

\affil[1]{\raggedright Department of Bioengineering}
\affil[2]{Department of Electrical and Computer Engineering}
\affil[3]{Department of Psychology}
\affil[4]{Department of Neurobiology}
\affil[5]{Neurosciences Graduate Program}
\affil[ ]{University of California, San Diego, 9500 Gilman Drive, La Jolla, CA 92093, USA}
\affil[$\dagger$]{These authors contributed equally to this work.}
\affil[$\blacklozenge$]{These authors jointly supervised this work.}

\begin{document}
\maketitle
\begin{abstract}
Complex, learned motor behaviors involve the coordination of large-scale neural activity
across multiple brain regions, but our understanding of the population-level dynamics
within different regions tied to the same behavior remains limited. Here, we investigate
the neural population dynamics underlying learned vocal production in awake-singing
songbirds. We use Neuropixels probes to record the simultaneous extracellular activity
of populations of neurons in two regions of the vocal motor pathway. In line with
observations made in non-human primates during limb-based motor tasks, we show that
the population-level activity in both the premotor nucleus HVC and the motor nucleus
RA is organized on low-dimensional neural manifolds upon which coordinated neural
activity is well described by temporally structured trajectories during singing behavior.
Both the HVC and RA latent trajectories provide relevant information to predict vocal
sequence transitions between song syllables. However, the dynamics of these latent
trajectories differ between regions. Our state-space models suggest a unique and
continuous-over-time correspondence between the latent space of RA and vocal output,
whereas the corresponding relationship for HVC exhibits a higher degree of neural
variability. We then demonstrate that comparable high-fidelity reconstruction of
continuous vocal outputs can be achieved from HVC and RA neural latents and spiking
activity. Unlike those that use spiking activity, however, decoding models using neural
latents generalize to novel sub-populations in each region, consistent with the existence
of preserved manifolds that confine vocal-motor activity in HVC and RA.
\end {abstract}


\begin{corrauthor}
tgentner@ucsd.edu (TQG), vgilja@ucsd.edu (VG)
\end{corrauthor}

\section*{Introduction}
Sensory, motor and cognitive processes require coordinated activity across large populations of neurons. While the study of cortical function, particularly in motor systems, has traditionally focused on relating single-unit activity to specific behavior covariates~\cite{georgopoulos_1982, evarts_1968, thach_1978, morrow_2007}, commensurate with our increasing capacity to record simultaneously from larger and larger neural populations, analysis  methods have moved beyond the search for representations at the single-neuron level~\cite{churchland_neural_2012, shenoy_cortical_2013, sussillo_neural_2015}. In the motor system, this has led to the theory of ‘neural manifolds for the control of movement’, which shifts away from the necessity of single neurons to explicitly represent movement covariates in favor of a paradigm where the collective activity of interconnected neurons underlies the computations associated with movement planning, execution and sensory integration~\cite{gallego_neural_2017, gallego_long-term_2020}. In this view, the spiking activity of individual neurons within such networks is regarded as a stochastic process driven by the underlying time-varying state of the network performing computations dynamically. This perspective has provided new insights on how brain activity relates to behavior across a wide range of motor tasks~\cite{gallego_neural_2017, gallego_cortical_2018}, the role of interconnected brain regions during behavior and learning~\cite{lara_different_2018, russo_neural_2020, perich_motor_2020}, and the mechanisms of information transfer across neural networks~\cite{semedo_cortical_2019} that were not evident from single-neuron or trial-averaged responses. The challenge in understanding network functionality arises from the substantial size of neural populations across various brain regions. However, the intrinsic interconnectivity within neurons in these populations suggests that the diversity of observable spiking patterns is inherently limited by the network's connectivity structure~\cite{okun_diverse_2015}. As a result, the population activity patterns are likely restricted to a neural space of lower dimensionality than the total number of neurons in the network. This idea implies that, despite the vast number of neurons within a given population, the effective computational space utilized by the network is expected to be constrained to a neural manifold characterized by a limited set of independent "neural modes" (latent dimensions). The temporal activation of these modes defines the time-evolution of the network’s dynamics during neural processing~\cite{vyas_computation_2020}. Neural decoding studies support this hypothesis through accurate prediction of behavioral variables across several widely-studied motor tasks, despite having access to only a small fraction of neurons within the network~\cite{barroso_decoding_2019, ethier_restoration_2012, carmena_learning_2003, sadtler_neural_2014}.  

Numerous studies have used dimensionality reduction methods to identify manifolds that effectively capture neural variability and uncover the population dynamics driving behavior on an individual trial basis. These approaches circumvent the necessity for trial-averaging or time-warping strategies to uncover neural organization, thereby preserving the natural neural variability inherent in each trial~\cite{vyas_computation_2020, pandarinath_inferring_2018, farshchian_adversarial_nodate, altan_estimating_2021, russo_motor_2018, degenhart_stabilization_2020, yu_gaussian-process_2009}. Although a significant portion of this research has centered on dorsal premotor (PMd) and primary motor (M1) population recordings during various motor control tasks, including movement planning and motor learning~\cite{gallego_cortical_2018, safaie_preserved_2022, schroeder_cortical_2022, sadtler_neural_2014, perich_neural_2018, vyas_neural_2018}, these techniques have been extended to study behavioral tasks and brain regions beyond motor cortices~\cite{schneider2023cebra, yu2022_calcium, kennedy_stimulus-specific_2020, thura_unified_2020}. One prevalent limitation in current research is the frequent emphasis on basic tasks performed in constrained environments, where movement is restricted to a few degrees of freedom. As such, several important questions remain. Can population activity tied to naturalistic behaviors of higher complexity be similarly characterized by dynamics within low-dimensional neural manifolds~\cite{suresh_neural_2020}? Can neural manifolds help reveal function at the population level in upstream brain areas that integrate higher-order neural inputs~\cite{russo_neural_2020}? And, how do the underlying dynamics of neural populations differ between brain regions driving behavior jointly~\cite{perich_neural_2018}? We address these questions in the zebra finch (Taeniopygia guttata) songbird model.
 
Vocal learning is a rare evolutionary trait found only in a limited number of mammalian and avian species~\cite{jarvis_evolution_2019, sainburg_parallels_2019}. Despite important differences with humans, songbirds have emerged as excellent model systems to study the neural bases of vocal learning, perception, and production~\cite{davenport_birdsong_2023, brainard_what_2002, cohen_hidden_2020, gentner_recursive_2006, brown_local_2021, mello_zebra_2014, gentner_neuronal_2003, arneodo_neurally_2021}. Birdsong, like human speech, is a learned vocal communication behavior critical in conspecific social interactions. During birdsong, adult male zebra finches produce renditions of moderately stereotyped motifs, defined as structured sequences of 3 to 10 distinct, short ($30$ $\unit{ms}$ to $60$ $\unit{ms}$) vocal elements (syllables), alternating with introductory notes, intra-motif notes and other vocalizations that together constitute a song bout. The precise coordination of muscles in the respiratory-vocal apparatus necessary to produce this complex behavior is orchestrated by the song control system. This vocal-motor neural circuitry is the primary driver of adult zebra finch song generation, and comprises the telencephalic regions HVC (used as a proper name) and robust nucleus of the arcopallium (RA)~\cite{hahnloser_ultra-sparse_2002, fee_neural_2004, leonardo_ensemble_2005}. HVC interneurons ($HVC_{int}$) are believed to play a role in motor control and auditory integration functions,  processing neural inputs from neighboring brain pathways to drive the propagation of vocal sequences~\cite{elmaleh_sleep_2021, moll_thalamus_2023, okubo_growth_2015, picardo_population-level_2016}. Conversely, HVC projection neurons ($HVC_{pn}$) have been observed to output precise bursts of activity to adjacent brain areas, providing a continuous and uniform representation throughout the song~\cite{lynch_rhythmic_2016}. Neuronal subtypes of $HVC_{pn}$ target RA, the basal ganglia nucleus Area X, and the auditory nucleus Avalanche~\cite{mooney_hvc_2005, rauske_state_2003, roberts_identification_2017}. RA efferents target brainstem motor neurons that directly innervate effector organs in the respiratory-vocal apparatus to produce song~\cite{fee_neural_2004}.

Traditional electrophysiological studies in both HVC and RA have focused on relating single-neuron responses, typically recorded serially, to behavior while assuming independence across neurons within the network~\cite{fee_neural_2004, hahnloser_ultra-sparse_2002, leonardo_ensemble_2005, kozhevnikov_singing-related_2007}. Even in studies that have recorded simultaneously from larger populations~\cite{picardo_population-level_2016, elmaleh_sleep_2021, lynch_rhythmic_2016}, the dynamics that emerge from patterns of covariation across neurons have not been directly assessed. Furthermore, in the study of HVC function, research has frequently centered around the $HVC_{pn}$ population activity and its role as a continuous-time code for song production and learning~\cite{picardo_population-level_2016, long_support_2010, fee_hypothesis_2011, long_temperature_2008, glaze_behavioral_2007, leonardo_ensemble_2005, hahnloser_ultra-sparse_2002, katlowitz_stable_2018}. As a consequence, the activity of the $HVC_{int}$ population has been less frequently explored, despite exhibiting song-locked firing rate modulations and being expected to play a significant role in sensory integration and motor planning~\cite{lynch_rhythmic_2016, markowitz_mesoscopic_2015, kosche_interplay_2015, kozhevnikov_singing-related_2007}. Here we set out to examine the latent dynamics that govern the spiking activity of HVC and RA neural populations in singing birds. The simultaneous recording of HVC and RA activity enables their direct comparison in equivalent experimental settings. We focus first on the intrinsic dynamics of HVC and RA during singing, testing the ability of our models to identify low-dimensional neural manifolds in networks with differing population sizes and spiking characteristics. We find that the population-level activity in both HVC and RA is well described by smooth, reproducible neural trajectories on low-dimensional latent manifolds that capture a significant amount of neural variance, and are robust to changes in population size and neural sub-types. In RA, we uncover highly reproducible, invariant neural dynamics during consistent singing, resembling dynamics in the motor cortex of non-human primates during forelimb motor tasks~\cite{churchland_neural_2012}. In HVC, we uncover neural dynamics that also exhibit a clear structure in their rhythmic oscillations during song, but manifest a larger degree of temporal variability than those in RA. We then show that the continuous latent neural trajectories in both HVC and RA carry sufficient information to predict timely transitions in the observed sequence of song syllables, consistent with the continuous-time coding hypothesis for HVC function~\cite{lynch_rhythmic_2016, elmaleh_sleep_2021}. Lastly, we demonstrate how knowledge of underlying latent manifolds in both HVC and RA allows a decoding model that synthesizes continuous vocalizations directly from neural dynamics to generalize across novel neural populations. 

\begin{figure*}
\centering
  \includegraphics[width=0.97\textwidth]{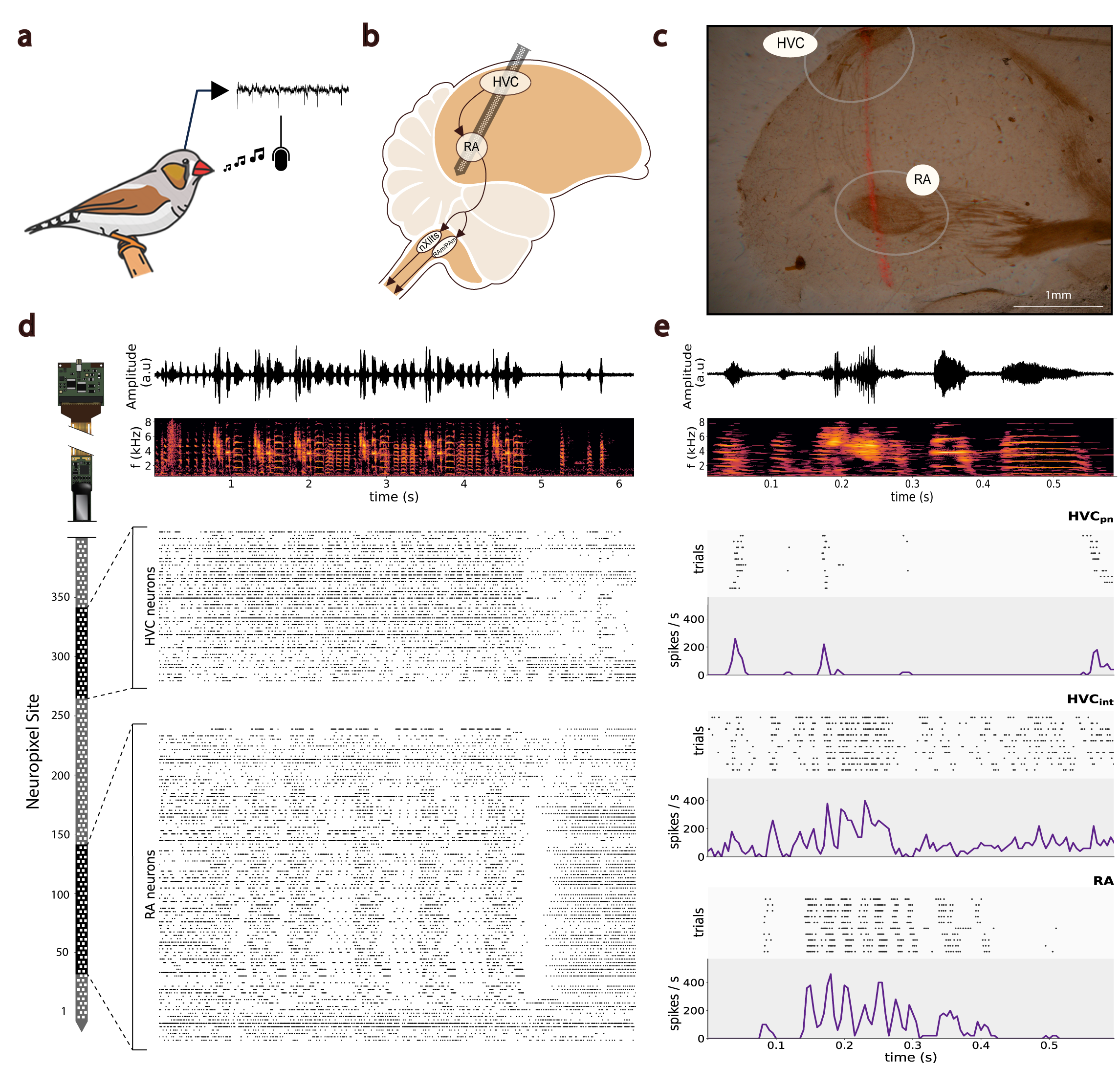}
  \caption{\textbf{Multi-Region Neuropixels Recordings in Awake-Singing Songbirds.} \textbf{a.} Illustration of the experimental objective to record neural activity alongside the song of freely-singing zebra finches. \textbf{b.} A single Neuropixels probe is implanted chronically at the required angle to simultaneously record the extracellular activity of HVC and RA neural populations. \textbf{c.} Example histological sample for confirmation of recording location, showing electrode track (fluorescent trace) through a saggittal section of the caudal telencephalon with both HVC and RA visible. \textbf{d.} Amplitude waveform and corresponding spectrogram of a recorded song bout (top). Below, spike-trains from spike-sorted single-unit activity (SUA) and multi-unit activity (MUA) clusters in the HVC and RA regions are time-aligned with the audio recording. On the left, a sketch of the Neuropixels probe illustrates the probe sites covering each brain region, estimated from the probe’s geometry, brain histology, and observed modulated neural activity. \textbf{e.} Amplitude waveform and spectrogram of an example motif within the bout (top). Below, example raster plots and PSTHs of one putative HVC projection neuron, one putative HVC interneuron, and one putative RA neuron across 10 renditions of the same motif and aligned to the onset of the first syllable in the motif. 
}
  \label{fig:fig1}
\end{figure*}

\section*{Results}
We hypothesized that the model of neural manifolds for the control of movement~\cite{gallego_neural_2017} extends to behaviors of increased complexity across various species, including vocal production in songbirds. To investigate the intrinsic neural dynamics driving the population activity of brain regions in the vocal-motor pathway (HVC and RA) during song, we implanted high-density Neuropixels silicon probes in zebra finches engaged in unconstrained vocal behavior (n = 4 birds) (\hyperref[fig:fig1]{Figure 1}; \hyperlink{methods:surgical}{Methods}). In each recording session, we aimed to capture the simultaneous activity of populations of neurons in the premotor region HVC and the motor region RA alongside the bird’s own song. We successfully recorded RA activity during awake-singing in all four birds (\hyperref[S1]{Sup. Fig. 1}). In bird A and bird B, we were able to record the activity of both the HVC and RA populations simultaneously. In bird C the implant yielded RA-only recordings. In bird D, we achieved partial targeting of HVC, but the neuron-count yield was insufficient to conduct population-based analyses (\hyperref[S2]{Sup. Fig. 2}). During each recording session, we identified neural clusters that were classified into single-unit activity (SUA) and multi-unit activity (MUA) (\hyperlink{methods:neural_processing}{Methods}). SUA clusters in HVC were further classified into putative HVC interneurons ($HVC_{int}$) and putative HVC projection neurons ($HVC_{pn}$) (\hyperref[fig:fig1]{Figure 1e}; \hyperlink{methods:neural_processing}{Methods}). Across all birds and sessions, we recorded an average of $82 \pm 2$ (mean $\pm$ s.d.; range, 80-85) electrode sites within RA and an average of $36 \pm 22$ (mean $\pm$ s.d.; range, 9-64) electrode sites within HVC. From those sites, we isolated an average of $105 \pm 45$ (mean $\pm$ s.d.; range, 57-179) putative RA and $59 \pm 47$ (mean $\pm$ s.d.; range, 7-121) putative HVC neural clusters that were stable throughout a given session. Aligning the spike-sorted activity from HVC and RA clusters with the song bouts revealed visually discernible modulations in firing rates that were time-locked to the onset/offset of the song in both populations. Across birds, a subpopulation of well-isolated RA neurons exhibited highly structured, coordinated patterns of activity temporally aligned to song motifs. We observed a visible drop in the spiking activity of both neural populations time-locked to the termination of the bout. This drop was particularly evident in RA subpopulations, which exhibited a brief period of fully suppressed neural activity (\hyperref[fig:fig1]{Figure 1d}).

\begin{figure*}
\centering
\includegraphics[width=0.97\textwidth]{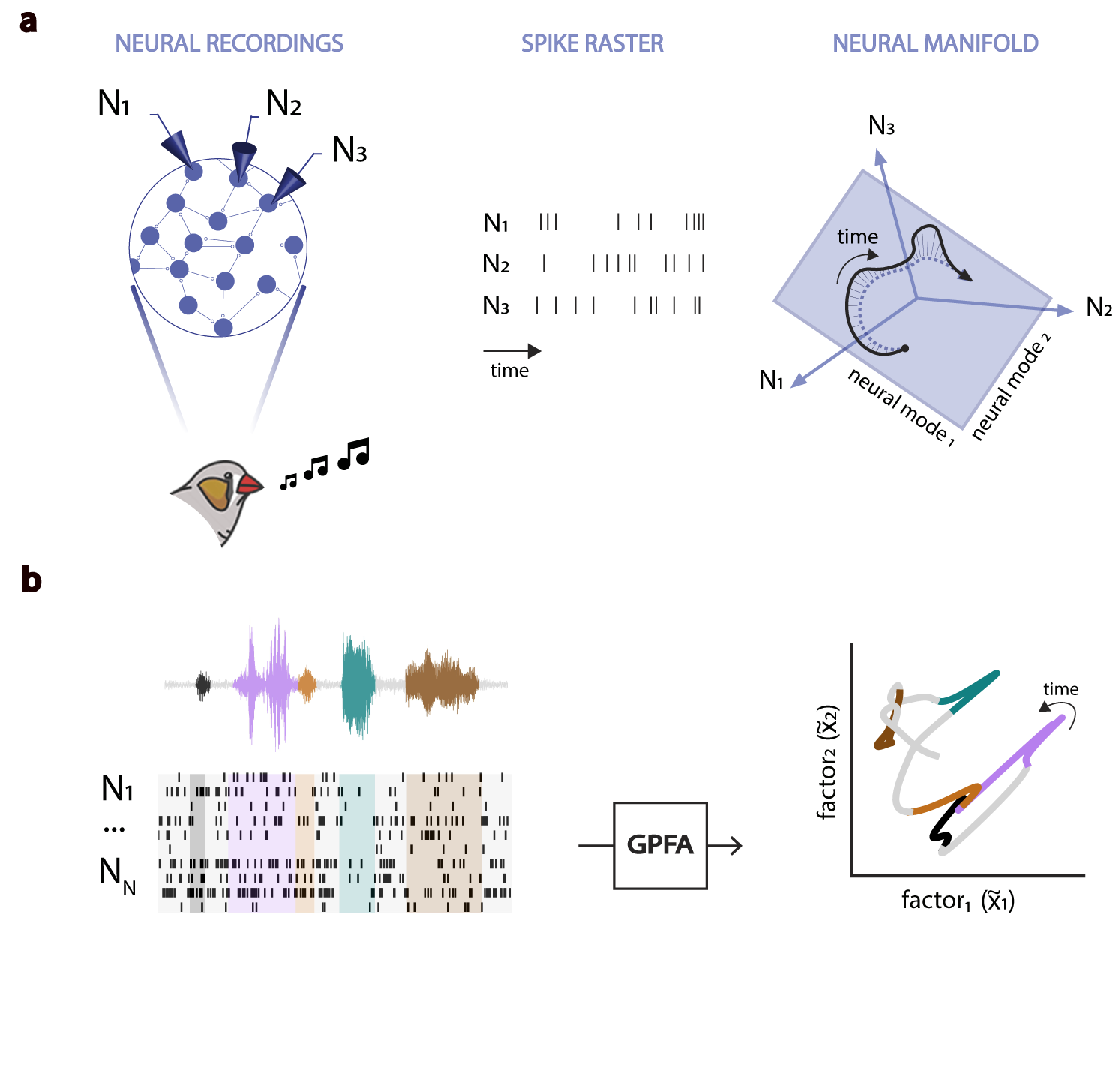}
  \vspace{0cm}
  \caption{\textbf{The Neural Manifold Hypothesis.} \textbf{a.} (Adapted from Gallego et al. 2017). Birdsong production involves coordinating muscle groups via the activity of interconnected neural populations in the song-motor pathway (HVC and RA). Despite our ability to sample only from a limited number of neurons within these brain regions, the observable patterns of population activity are expected to be restricted by their intrinsic connectivity. The time-varying population activity can be described by a time-dependent latent trajectory of network states confined to a low-dimensional neural manifold constrained by the network connectivity. The combined activity of the neural modes defining such low-dimensional manifold explains the specific activity of individual neurons within the population. \textbf{b.} Gaussian-Process Factor Analysis (GPFA) is utilized to estimate low-dimensional neural manifolds that capture the intrinsic covariance structure of the coordinated spiking activity of neural populations. We examine the latent trajectories described by the activity of the HVC and RA populations in such manifolds to ask critical questions about how the brain orchestrates the execution of complex vocal behavior. Are the dynamics of these latent trajectories informative of time-dependent changes in the observed vocalizations? Do the latent manifolds of different brain regions involved in the execution of the same task exhibit similar characteristics? How do the underlying dynamics of neural populations differ between brain regions driving behavior jointly?
}
  \label{fig:fig2}
\end{figure*}

\begin{figure*}
\centering
\includegraphics[width=0.97\textwidth]{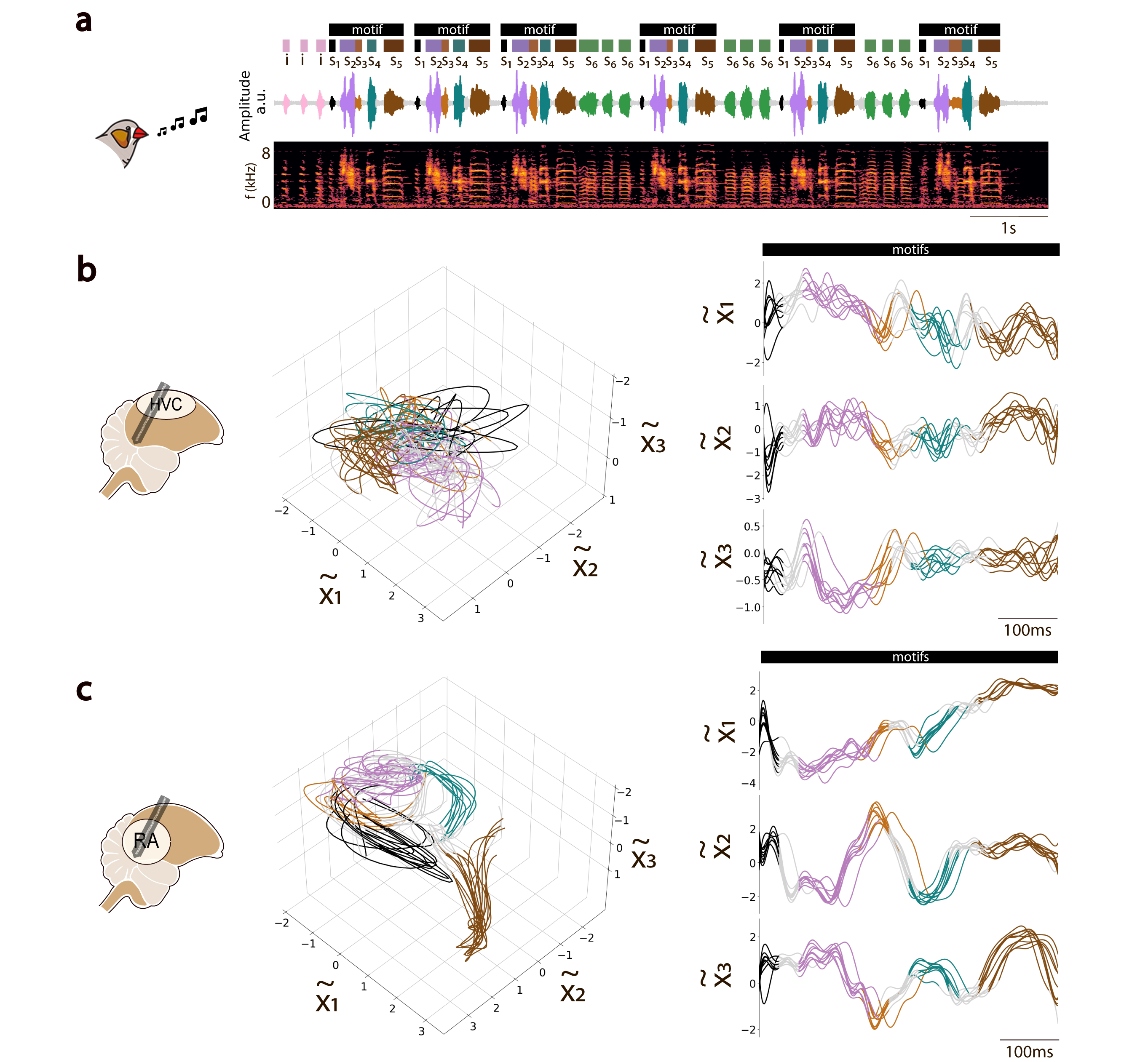}
  \caption{\textbf{State-Space Analysis of Motor Nuclei Activity in the Song-Motor Pathway.} \textbf{a.} Amplitude waveform and spectrogram of a recorded song bout from a male zebra finch. The bout features multiple motifs, comprising individual syllables ($S_1$ to $S_5$), color-coded from black to brown, with additional intra-motif notes ($S_6$, green) and introductory notes ($i$, pink) preceding the first motif. \textbf{b.} HVC neural trajectories during motif repetition, inferred by projecting HVC activity onto a latent manifold defined by the leading three GPFA factors (left). Trajectory subsegments are color-coded to match associated song syllables. The same trajectories are also shown projected onto each of the three-dimensional axes (GPFA factors), arranged in descending order by variance explained (right). \textbf{c.} Same as in (b) but for GPFA-inferred neural trajectories in RA.  Comparing b and c highlights the larger degree of smoothness and temporal structure in the activity of RA across vocal repetitions compared to the greater variability observed in HVC during the same song bout.
}
  \label{fig:fig3}
\end{figure*}

To better understand governing dynamics of HVC and RA populations during vocal behavior, we applied Gaussian-Processes Factor Analysis (GPFA) to identify low-dimensional neural manifolds that capture the shared neural variability in a given population~\cite{yu_gaussian-process_2009} (\hyperref[fig:fig2]{Figure 2}; \hyperlink{methods:dynamics}{Methods}). Examples of 3-dimensional manifolds defined by the three leading neural modes (GPFA factors) ($\tilde{x}_{1}-\tilde{x}_{3}$) that capture the greatest degree of spiking covariation during song are shown in \hyperref[fig:fig3]{Figure 3}, where moment-to-moment fluctuations in population activity appear as visually interpretable trajectories on the manifold. The RA population activity described highly-stereotyped, low-dimensional trajectories during renditions of song motifs with similar spectral characteristics (\hyperref[fig:fig3]{Figure 3b}). In contrast, latent trajectories of HVC populations were much more variable, but nonetheless preserved a high-degree of temporal structure, as illustrated by the 1D-trajectories shown in \hyperref[fig:fig3]{Figure 3c}. In both HVC and RA, we observed clearly distinct neural state-regions (localized regions within the manifold) associated with the production of different motif syllables (\hyperref[fig:fig3]{Figure 3b-c}). 

\begin{figure*}
\centering
\includegraphics[width=0.97\textwidth]{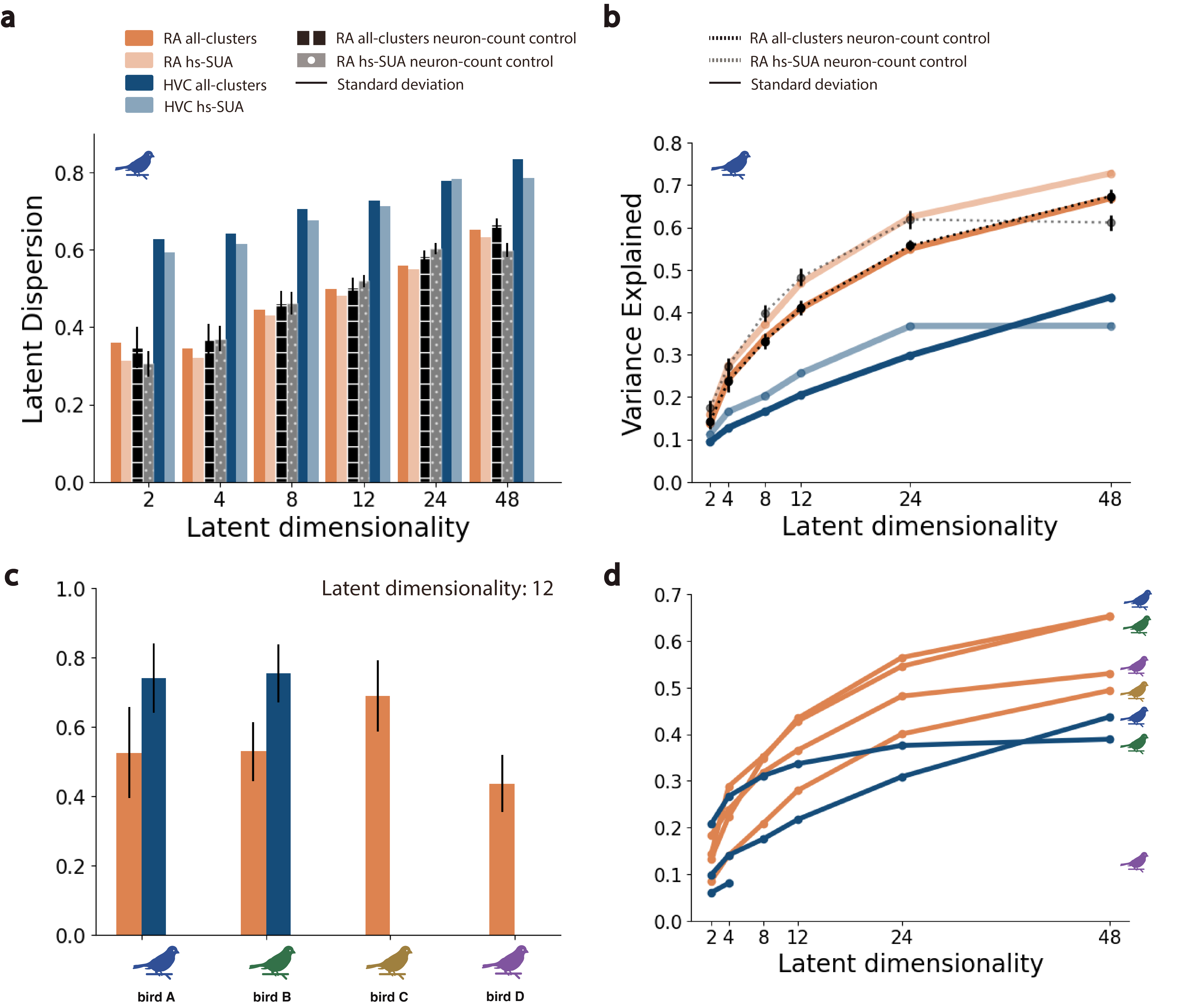}
  \caption{\textbf{Dimensionality-Dependent Analysis of Latent Neural Manifolds.} \textbf{a.} Mean latent dispersion, calculated as the average deviation from the mean trajectory, observed in neural manifolds inferred through GPFA as a function of latent dimensionality across different neural populations in a selected bird. Two distinct neural populations corresponding to well-isolated, high-spiking single-units (hs-SUA) and to a combination of all isolated SUA and MUA clusters (all-clusters) were considered within HVC and RA recordings. The latent dispersion for by the inferred manifolds remains stable, irrespective of the neuron type or the sorting quality within a subpopulation. Controls randomly subsampling from the RA population to match the sizes of the HVC and RA populations confirm that the resulting manifolds were robust to variations in population sizes in our recordings. \textbf{b.} Neural variance captured by manifolds inferred through GPFA as a function of latent dimensionality across different neural populations in a selected bird. The variance accounted for by the inferred manifolds remains consistent, irrespective of the neuron type or the sorting quality within a subpopulation. \textbf{c.} Latent dispersion observed in 12-dimensional latent manifolds across four different songbird individuals. \textbf{d.} Neural variance captured by latent manifolds across four different songbird individuals.
}
  \label{fig:fig4}
\end{figure*}

Qualitatively, the low-dimensional RA latent trajectories appeared to be less variable than those of HVC (\hyperref[fig:fig3]{Figure 3b,c}). To quantify this observation, we computed the \textit{latent dispersion} of trajectories on the manifold, defined as the normalized average deviation from the mean trajectory (\hyperlink{methods:dynamics}{Methods}). The latent dispersion of HVC trajectories was consistently larger than that for RA trajectories (ANOVA, Tukey's HSD test, $p-value < 0.05$; \hyperref[fig:fig4]{Figure 4a,c}; \hyperlink{methods:controls}{Methods}). To understand whether this effect might be tied to physiological sampling biases between regions, we computed the dispersion for GPFA-inferred latent trajectories in several control conditions (\hyperlink{methods:controls}{Methods}). These address potential effects due to different population sample sizes between HVC and RA, differences in the average spike rate (and total spike count) between HVC and RA, the presence or absence of sparse-firing projection neurons in HVC, and single-unit isolation quality. Whereas there was not a significant difference in the latent dispersion measurements between most of the whithin-region (HVC or RA) seleted subpopulations (ANOVA, Tukey's HSD test, $p-value > 0.1$), in all cases, the latent dispersion remained significantly higher in RA than in HVC (ANOVA, Tukey's HSD test, $p-value < 0.001$; \hyperref[S3]{Sup. Fig. 3}; \hyperref[S4]{Sup. Fig. 4}; \hyperref[S5]{Sup. Fig. 5}; \hyperref[S6]{Sup. Fig. 6}; \hyperlink{methods:controls}{Methods}). This effect was consistent across multiple populations, and for manifolds of different dimensions (\hyperref[fig:fig4]{Figure 4}). 

We also considered how well the GPFA-inferred manifolds captured the empirical response variance by computing the amount of population activity variance explained by the inferred latent manifolds, as a function of manifold dimensionality (\hyperref[fig:fig4]{Figure 4}). For both HVC and RA, the manifolds at all tested dimensionalities (from 2 to 48) explained significantly more of the empirical response variance than that expected by chance (\hyperref[S3]{Sup. Fig. 3}; \hyperref[S4]{Sup. Fig. 4}; \hyperref[S5]{Sup. Fig. 5}; \hyperref[S6]{Sup. Fig. 6}). In nearly all cases across all birds, the inferred latent manifolds for RA explained a consistently larger amount of variability compared to those for HVC (ANOVA, Tukey's HSD test, $p-value < 0.001$; \hyperref[fig:fig4]{Figure 4}). The upper bound on explained variance is not clear given current data limits, but the observed trends suggest an asymptotic plateau after 24-dimensional neural manifolds for the studied vocal task (\hyperref[fig:fig4]{Figure 4b,d}). We note that an asymptote for the explained variance below 1.0 is expected, because unlike PCA, factor analysis models only the covariance in the input variables (in our case neural responses) and attributes remaining response variance to private (anisotropic) noise within each neuron. Thus, the differing asymptotes between regions suggest that a larger proportion of the total response variance in HVC is attributable to this private variability compared to RA. This private variance may be related to differences in the heterogeneity of HVC inputs and/or outputs compared to RA, or differences in neuronal response profiles. To test the possibility that the GPFA upper bound may be related directly to single neuron response characteristics, we identified two different HVC and RA sub-populations (\hyperlink{methods:controls}{Methods}). The first sub-population, termed \textit{hs-SUA}, included only well-isolated SUA clusters exhibiting high-spiking characteristics ($>$ 10Hz on average across vocal renditions). In HVC, the hs-SUA sub-population excluded most of the putative $HVC_{pn}$, characterized by their sparse, bursty activity, which might bias our manifold models ($93\%$ and $85\%$ putative $HVC_{pn}$ exclusion in bird's A and B respectively). The second sub-population, termed \textit{all-clusters}, included all high-quality SUA and MUA cluster identified in either population (\hyperlink{methods:controls}{Methods}). The variance explained and latent dispersion metrics yielded consistent results independently of the neural sub-population selected (\hyperref[fig:fig4]{Figure 4}; \hyperlink{methods:controls}{Methods}).

\begin{figure*}
\centering
\includegraphics[width=0.97\textwidth]{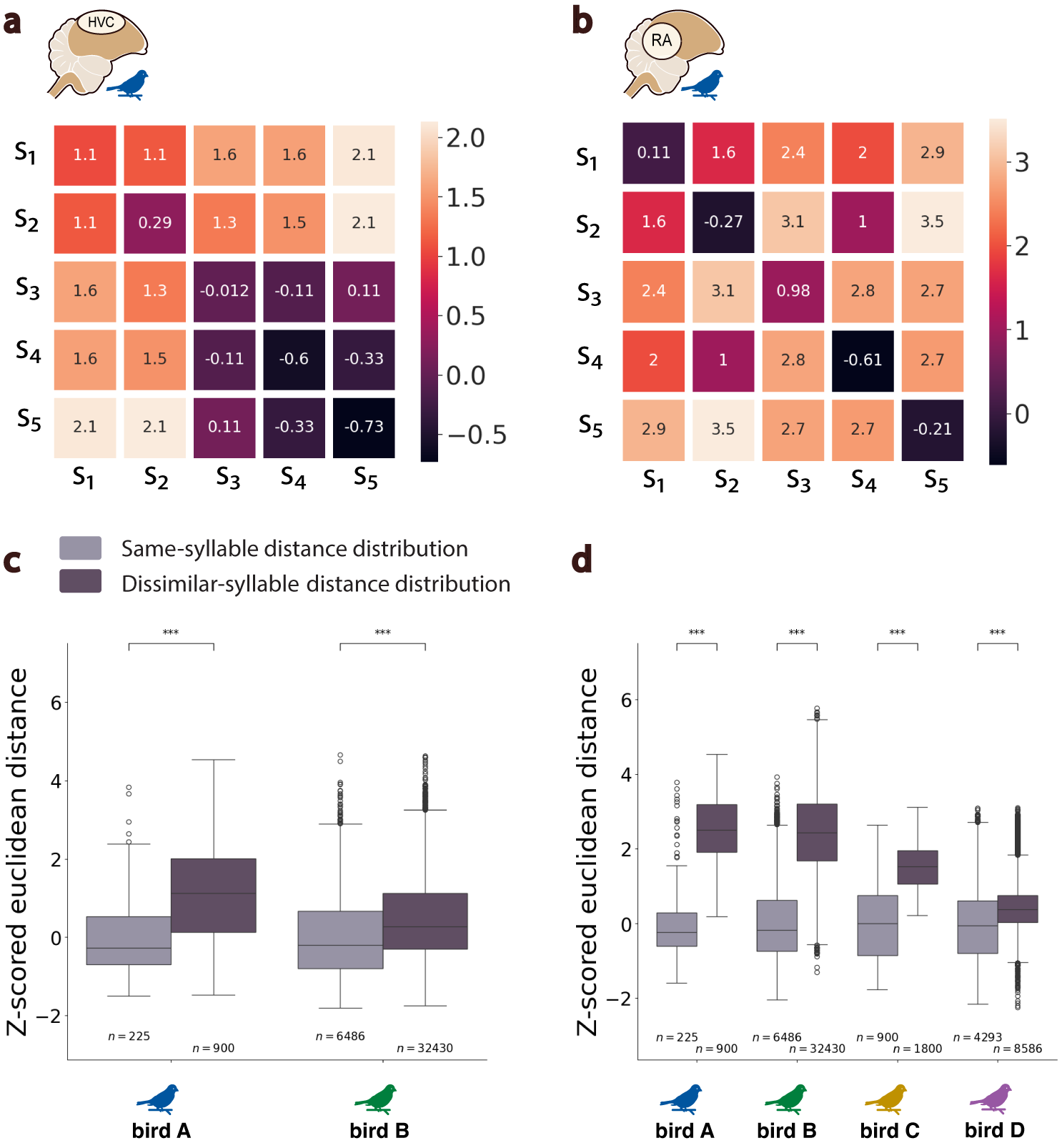}
  \caption{\textbf{Latent Manifold States vary by Syllable Distances.}
 \textbf{a.} Confusion matrix showing normalized, mean Euclidean distances across neural states corresponding to the production of different syllables (S$_1$ to S$_5$) in bird A in \textbf{a.} HVC latent manifold and \textbf{b.} RA latent manifold. \textbf{c.} Boxplots showing distributions in two birds of z-score normalized distances between HVC neural latent states that correspond to vocalization of either the same (light gray) or dissimilar syllables (dark gray) across motif renditions. Values are z-scored relative to states corresponding to the same syllables in a given bird. \textbf{d.} AS in (c) but for distributions of distances across RA neural latent states. *** denote $p-value < 0.001$ (Welch's t-test).
}
  \label{fig:fig5}
\end{figure*}

We next considered the relationship between the latent neural dynamics in each region and vocal behavior. We first quantified the extent to which neural states associated with the production of different (or similar) discrete vocal elements (syllables) were separable (or not) within an inferred neural manifold. To do this, we parse neural trajectories into "states" based on syllable onsets and offsets derived from behavior, then compared the distribution of pairwise distances between neural state trajectories associated with the production of either the same or different song syllables (\hyperref[fig:fig5]{Figure 5a,b}; \hyperref[S7]{Sup. Fig. 7}; \hyperlink{methods:distance}{Methods}). For manifolds in both HVC and RA, the pairwise differences between different syllables were significantly greater than differences between similar syllables (Welch’s t-test, $p-value < 0.001$; \hyperref[fig:fig5]{Figure 5}). Thus, neural trajectories on nearby regions of the manifold tend to be associated with the production of similar vocalizations.  Consistent with the dispersion and overall variability in the manifolds from HVC and RA, the separability of neural latent states between different syllable types was significantly stronger for RA neural trajectories compared to HVC (Wilcoxon signed-rank test, $p-value < 0.001$; \hyperref[fig:fig5]{Figure 5c,d}).

\begin{figure*}
\centering  
\includegraphics[width=0.8\textwidth]{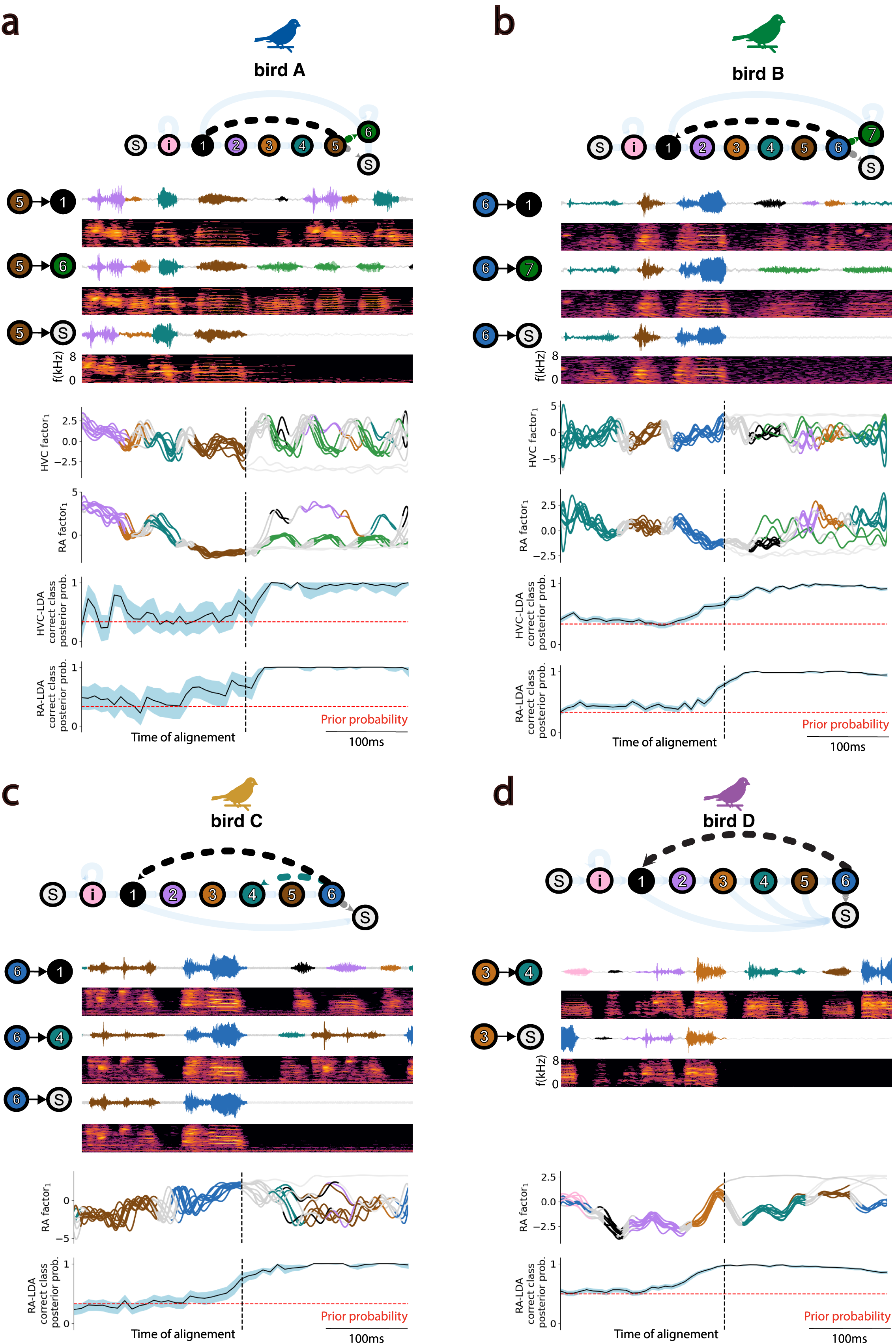}
  \caption{\textbf{HVC and RA Latent States predict Vocal Transitions.} Each panel (a-d) features a state diagram at the top illustrating both deterministic and probabilistic vocal transitions between syllables (colored circles) in the song of individual birds. Beneath, amplitude waveforms and corresponding spectrograms display examples of specific probabilistic transitions following a selected \textit{branching syllable} in each bird’s own song. Below, HVC and/or RA neural trajectories for ten randomly chosen vocal renditions are aligned to the end of the \textit{branching syllable} and projected onto the leading GPFA factor. At the bottom of each panel, results from Linear Discriminant Analysis (LDA) showing instantaneous posterior (and prior, red) probabilities of accurately predicting the vocal transition class at each time-point from the corresponding HVC and/or RA latent state. The $95\%$ confidence intervals are shown around mean posterior probabilities.
}
  \label{fig:fig6}
\end{figure*}

The foregoing analysis captures temporal structure in the neural trajectories corresponding to within-syllable acoustics, but misses longer time scales that might relate to the sequencing of syllables across motifs. To investigate relationships between latent neural dynamics and the sequential structure of vocal behavior, we asked whether the transitions between syllables could be predicted from their associated HVC and/or RA latent neural states (\hyperlink{methods:branching}{Methods}). Because many syllable transitions in zebra finch song are deterministic, we first identified “branch points” in each birdsong where the syllable sequencing was probabilistic. The variability in syllable sequencing typical of Zebra finch song is illustrated in the behavioral state diagrams shown in (\hyperref[fig:fig6]{Figure 6}). A representative example is bird A, presenting 7 different song syllables (syllables 1-6 and i) and 3 “branch points” (syllables i, 5 and 6) where sequencing is probabilistic rather than deterministic. For example, syllable 5 may be followed by syllables 1, 6 or silence. We annotated motif renditions based on the type of transition that occurred after the highlighted “branch point” and asked whether we could predict the occurring vocal transition type from the sequence of HVC and RA latent states. We trained a Fisher's linear discriminant analysis (LDA) to estimate the posterior probability of the correct vocal rendition  type from neural states at each time point along the latent trajectory. We used a bootstrap analysis (1000 runs) and evaluated the model's performance using a leave-one-out cross-validation scheme in each run (\hyperlink{methods:branching}{Methods}). As shown in \hyperref[fig:fig6]{Figure 6}, we were able to predict probabilistic transitions in the sequence of song syllables with high reliability and temporal precision from both HVC and RA latent neural states. Thus, both the HVC and RA neural latents carry relevant temporal information about syllable transitions within the song. 

\begin{figure*}
\centering
\includegraphics[width=0.8\textwidth]{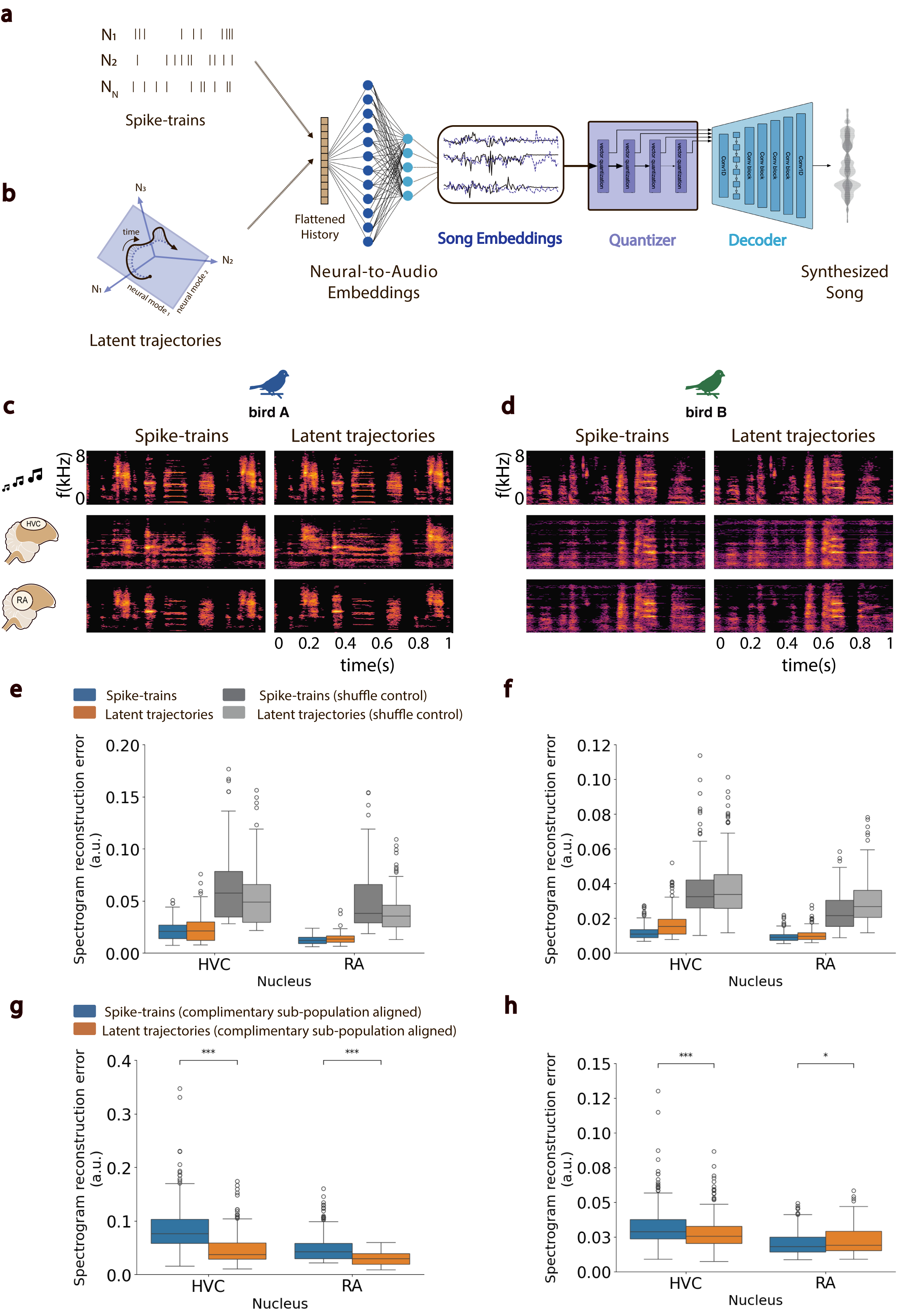}
  \caption{\textbf{Brain-to-Song Decoder.} \textbf{a-b.} Architecture of the proposed decoder featuring a Feed-Forward Neural Network that predicts song embeddings from spike-trains (a) or neural latents (b), integrated with Quantizer-Decoder networks from a pre-trained EnCodec model. \textbf{c-d.} Spectrogram of a single-second sample of a bird’s own song (BOS, top row), compared with songs synthesized from HVC spike-trains (middle left), HVC neural latents (middle right), RA spike-trains (lower left), and RA neural latents (lower right) for two distinct birds (c) and (d). \textbf{e-f.} Boxplots depicting the mean squared error in spectrogram reconstruction across synthesized song examples from spike-trains (blue) and neural latents (orange), including shuffle controls for the birds shown in (c) and (d). \textbf{g-h.} Latent stability analysis results. Boxplots show the mean squared error in spectrogram reconstruction of songs synthesized using a model trained on a randomly-selected half of a neural population and evaluated using the complementary half, aligned with the original using Canonical Correlation Analysis. *** denote $p-value < 0.001$; * denote $p-value < 0.05$ (Student's t-test).
}
  \label{fig:fig7}
\end{figure*}

Having demonstrated a correspondence between latent trajectories and song acoustics within and between syllable, we next examined the relationship between HVC and RA neural latent trajectories and continuous vocal behavior. We developed a novel neural decoder architecture, then tested its efficacy in reconstructing moment-to-moment vocalizations from neural activity. The decoder (\hyperref[fig:fig7]{Figure 7}) uses a feed-forward neural network (FFNN) to map neural activity onto embedded representations of contemporaneous song generated by a separate Encoder-Quantizer-Decoder network (\hyperlink{methods:training}{Methods}) derived from a pre-trained audio codec~\cite{defossez2022highfi}. The resulting FFNN-Quantizer-Decoder model, EnSongdec, is trained to predict new song renditions from novel neural activity. Unlike most brain-to-speech decoders which use a classification-based approach to predict transitions between abstract vocal elements~\cite{willett_high-performance_2023, Metzger2023-lk}, we decoded brain activity into continuous acoustic waveforms. We trained separate EnSongdec networks on either the recorded spike-trains or the GPFA-inferred latent trajectories from HVC and RA (\hyperref[fig:fig7]{Figure 7a,b}), and evaluated the performance of each decoder on reconstruction of held-out song motifs (\hyperlink{methods:training}{Methods}). Overall, the EnSongdec model performs remarkably well, yielding high-fidelity synthetic waveforms that very closely match the original vocalizations (\hyperref[fig:fig7]{Figure 7c,d}). Decoder accuracy, taken as the pixel-wise mean square error between original and reconstructed song spectrograms, was comparable for both decoding strategies (spike-trains and GPFA trajectories) and decoders trained on either HVC or RA activity performed well above chance in both birds A and B (Student’s t-test, $p-value < 0.001$; \hyperref[fig:fig7]{Figure 7e,f})). Consistent with our other results, RA-based decoders showed superior performance compared to those based on HVC activity. 

The EnSongdec model allows us to test a fundamental conclusion of this work: namely, that population activity within both HVC and RA during vocalization is confined to a stable low-dimensional manifold. This implies that, given the population activity of a sufficient subset of neurons in a given region, it is possible to infer the structure of the manifold that constrains the collective activity of all neurons in that region~\cite{pei2022neural}. If true, then a decoder trained on the latent trajectories from a subset of neurons in a region should generalize to latent trajectories from novel sub-populations of neurons in the same region. In contrast, a spike-train-based decoder should struggle to generalize across novel sub-populations in the same region since the direct mapping from spiking activity to stimulus is much less constrained. To test these ideas, we randomly selected half of the recorded neurons from our largest HVC and RA populations and trained decoders utilizing either spike-trains or GPFA-inferred latent trajectories. We then evaluated these decoders' performance using the remaining half of the neural population as the input signal. We used Canonical Correlation Analysis to align the inferred variables from both halves of the population, represented as either spike-trains or latent trajectories~\cite{canonical_2003, gallego_cortical_2018, sussillo_neural_2015, farshchian_adversarial_nodate}. Despite an expected decline in the overall quality of the synthesized song, in three out of the four pairwise comparisons across HVC and RA in two different birds, decoders trained on neural trajectories generalized to novel sub-populations significantly better than those trained on spike-trains (Student’s t-test, $p-value < 0.001$; \hyperref[fig:fig7]{Figure 7g,h}). This supports the notion that common latent manifolds underlie vocal-motor activity in both HVC and RA.

\section*{Discussion}
As recording capacities increase, it is becoming clear that widespread correlations in population dynamics tied to behavioral state are present across the forebrain~\cite{stringer_brainwide_2019}. As a consequence, population-level dynamics have been linked with increasing frequency to various cognitive phenomena and behavioral tasks~\cite{vyas_computation_2020, gallego_neural_2017, lara_different_2018, perich_motor_2020, russo_neural_2020}. Here, we related neural population dynamics to the complex learned vocalizations in a species of songbird, zebra finches. We developed the capability to record simultaneously from populations of neurons in the sensori-motor and pre-motor regions HVC and RA, respectively, in singing zebra finches. These recordings enable direct comparison of low-dimensional temporal dynamics within these two brain regions, whose responses during production of learned vocal signals have been well-studied at the single-neuron level. The stereotypy present in zebra finch song, coupled with our state-space analysis, allowed us to characterize behaviorally relevant population dynamics during periods of consistent behavior on a single-trial basis. The population-level activity in both HVC and RA is well-described by smooth, time-varying trajectories that lie on low-dimensional manifolds. In both regions, these trajectories correspond in multiple ways to learned vocal behavior (i.e., the bird's song).  Compared to RA, the latent trajectories of HVC population activity are more variable during awake-singing. Despite differences in variability, however, the latent states inferred from both HVC and RA population activity carry sufficient information to predict song syllable identities within bouts and precise transitions between song syllables. Trajectories in both regions can be reliably decoded to generate synthetic, continuous, and high fidelity reconstructions of vocal acoustic waveforms. Our findings, parallel observations of low-dimensional dynamics observed in M1 during reach-like motor activities in non-human primates~\cite{churchland_neural_2012, rouse_condition-dependent_2018, pandarinath_neural_2015}, and extend this dynamical framework to encompass mulitple brain regions underlying seemingly more complex learned motor actions. 

While at the surface level singing in birds and reaching in primates may appear to differ dramatically in their complexity, the neural substrates driving these motor behaviors share clear characteristics. Like the brainstem projections from mammalian primary motor cortex~\cite{Munoz-casteneda_MOp_2021}, RA also targets brainstem motor neurons, specifically retroambigualis and parambigualis (RAm/PAm) and the tracheo-syringeal part of the hypoglossal nucleus (nXIIts) that innervate effector muscles in the respiratory apparatus and the syrinx, respectively,~\cite{roberts_telencephalic_2008, wild_neural_2000}. We've shown that then spiking activity in both HVC and RA can be modeled as a dynamical system, where the current state of the neural population is primarily influenced by its previous states, similar to the proposed role of M1 as an autonomous pattern generator driving muscles that give rise to movement~\cite{sussillo_neural_2015, shenoy_cortical_2013, russo_motor_2018}. Yet, despite the widely reported spiking reliability of $HVC_{RA}$ projection neurons during birdsong~\cite{lynch_rhythmic_2016, long_support_2010, prather_precise_2008, kozhevnikov_singing-related_2007}, and multiple experiments demonstrating that $HVC_{int}$ firing rate modulations are time-locked to syllable onset/offset —suggesting a critical role in song initiation and production~\cite{lynch_rhythmic_2016, kozhevnikov_singing-related_2007}—the latent dynamics of the HVC population activity were significantly more variable than those of RA. The HVC population dynamics exhibited a higher degree of latent dispersion and explained less of the total response variance and covariance comparatively. These statistical properties of the singing-dependent latent trajectories in HVC indicate a less constrained correspondence between neural and behavioral state compared to RA. In line with this observation, a recent study reported that RA recordings provided a more precise means of estimating song-related spontaneous replay events than those in HVC despite having recorded populations of similar sizes in each region~\cite{elmaleh_sleep_2021}. Although the precise function of HVC interneuron activity is not fully understood, it is reasonable to hypothesize that subpopulations of HVC interneurons are tied to independent higher-order aspects of vocal behavior, including sensory input and output processing, manifested in in the trial-to-trial spiking variability observed in the $HVC_{int}$ population response (\hyperref[fig:fig1]{Figure 1e}). Different subnetworks within HVC would then exhibit independent covariances. In fact, various studies suggest HVC resembles Broca's area/premotor cortex in primates, both in function and genetic profile~\cite{jarvis_evolution_2019, davenport_birdsong_2023, jarvis_2014_human_bird, doupe_99_bird_human}. HVC neurons project directly onto RA, and indirectly through the anterior forebrain pathway (AFP) via the basal ganglia nucleus Area X~\cite{mooney_neurobiology_2009, roberts_identification_2017, hamaguchi_recurrent_2012}. Additionally, HVC receives neural projections from the auditory nucleus Avalanche, the sensorimotor nucleus interfacialis of the nidopallium (NIf), known to serve as the major auditory afferent of HVC~\cite{Piristine2016ASA, lewandowski_at_2013, coleman_synaptic_2004}, and the thalamic nucleus Uvaformis (Uva), essential for learned song production~\cite{moll_thalamus_2023}. From a dynamical systems perspective, given the numerous recurrent projections to/from adjacent brain areas, one could expect that HVC activity is mixed-driven by both afferent inputs and intrinsic dynamics. Consequently, methods that focus on maximizing the mutual information between related neural and sensory/behavioral latent spaces, along with those that systematically target afferent modulation of HVC dynamics, could prove useful for revealing neural manifolds associated with independent neural processes in this important brain region ~\cite{schneider2023cebra, Perkins_2023}.

It is possible that organized, untangled intrinsic dynamics exist in HVC, but in a higher dimensional space than in RA. Directly testing this hypothesis requires access to a larger neural sample that we obtained here, though we note that the unsupervised methods employed in this study presuppose that behaviorally-relevant dynamics will manifest within neural manifolds capturing the highest amount of covariance in the population activity. Nonetheless, it remains possible that structured, behaviorally-relevant, dynamics may reside within latent dimensions accounting for lesser fractions of neural variability, which could be the case if much of the recorded most appropriately attributable to song-related neural processes but not vocal acoustics, per se. In addition, we cannot fully rule out the possibility that the inherent complexity and potentially nonlinear nature of the intrinsic HVC dynamics, as well as the transformation from sampled high-dimensional neural spaces to latent neural manifolds, might exceed the analytical capacity of the current methodologies. That is, more powerful models may yet reveal better organized HVC dynamics. Nonetheless, the linear structure of GPFA has advantages in allowing for the separation of total response variance from neuron-to-neuron covariance, and from this perspective our findings show a significantly stronger, topographic correspondence between the neural latent space of RA and the output vocal space when compared to that of the HVC latent space. These differences are noteworthy as the specific roles of HVC during song production are still a matter of study~\cite{hamaguchi_distributed_2016, long_temperature_2008, andalman_control_2011}. 

Lastly, we investigated the continuous synthesis of zebra finch song from neural latents. A significant challenge in current neurally-driven communication prostheses lies in their ability to translate neural activity into sequences of words or phonemes, often at the cost of losing the prosodic elements in the reconstructed speech~\cite{willett_high-performance_2023, Metzger2023-lk}. We proposed a novel decoder design that translates neural activity into an embedded representation of time varying signals to then synthesize song in a continuous fashion. This architecture enabled reliable reconstruction of song spectrograms from both raw spike-trains and neural latent trajectories. Although high-quality song synthesis was possible using activity from both brain regions, decoders trained on RA activity produced reconstructions of higher fidelity compared to those trained on HVC activity. Interestingly, the performance of decoders trained on neural latent trajectories was comparable to those trained directly on spike-trains. Biologically, this supports our our central hypothesis that the behaviorally-relevant activity is evinced by the covariance structure of the population. For engineering applications, this result suggests the potential superiority of the dynamical approach in environments where neural stability is compromised, such as in the presence of acute temporal neural drift. To simulate such scenarios, we conducted experiments where decoders were trained and tested using disjoint subsamples from the HVC and RA recorded populations. We find that once the latent decoder is trained, it remains remarkably adept at decoding the activity of unique neural sub-populations for each region. Thus the covariance structure that defines the underlying manifold is a common attribute of the region, not of sub-samples from the region.  Decoding from neural latents, therefore, offers enhanced robustness against neuron turnover, highlighting its potential as a superior strategy in dynamic neural environments. We note that the lightweight computational complexity and real-time inference capabilities of the proposed architectures, coupled to the streamable features of EnCodec, meet the critical low-latency requirements essential for effective neurally-driven vocal prostheses. Future research should investigate the generalizability of these approaches across diverse vocal behaviors and more extensive vocal repertoires, with the goal of developing real-time brain-to-speech prostheses aiming to directly synthesize vocal output from neural signals encoding intent.

Collectively, our findings provide novel, population-based evidence on the differing neural computations performed by premotor region HVC and motor region RA during song, and leave an open question about the roles of HVC in integrating inputs from neighboring neural circuits and transmitting processed information to downstream brain pathways during sensorimotor control of learned vocal production. We posit that the manifold-based perspective of population activity in the avian brain sheds light on unrecognized but fundamental organizational features of different brain networks associated with the execution of complex vocalizations~\cite{greene_why_2023, perich_rethinking_2020, semedo_cortical_2019}. A deeper understanding of population-level neural interactions within and between regions in the song-motor pathway of the zebra finch and songbirds with higher-complexity song syntax can reveal valuable insights into the neural mechanisms underlying vocal production. More broadly, our work is motivated by recent progress towards the development of neurally-driven speech prostheses, which is considered a particularly challenging problem~\cite{sahin_sequential_2009, bouchard_control_2014, bouchard_functional_2013} due to the risks linked to the implantation of chronic electrode arrays in brain areas associated with this human behavior~\cite{martin_decoding_2018, bouchard_neural_2014, rabbani_potential_2019}. Whether and how these findings transfer to other vocal learners, including humans, remains unexplored. Regardless, deeper insight into the neural basis of vocal behavior will be essential for continued development of brain-machine interfaces that fully restore communicative abilities~\cite{willett_high-performance_2021, willett_high-performance_2023, teja_2016_speech, willett_hand_2020}. Our results reinforce the zebra finch, in particular, and the songbird, in general, as a functional animal proxy model to study auditory-motor control of learned vocal production.

\section*{Materials and methods}
All procedures were conducted in accordance with protocol S15027 approved and supervised by the University of California, San Diego Institutional Animal Care and Use Committee, and consistent with guidelines in the Guide for the Care and Use of Laboratory Animals.

\subsection*{\hypertarget{methods:habituation}{Habituation of zebra finches to acoustically-isolated environments}}
Prior to surgery, adult ($> 90$ days old) male zebra finches (Taeniopygia guttata), weighing between 16 \si{\g} and 20 \si{\g}, were acclimated to individual, acoustically-isolated recording chambers and held on a 14:10h light:dark cycle. Recording chambers were instrumented with cool-white LED strips (Adafruit Industries) for lighting, brushless DC motor (BDCM) cooling fans (Gdstime Technology Co.) for ventilation, an HP L6010 10.4-inch monitor (60 \si{\hertz} refresh rate, 4:3 aspect ratio, 1024 x 768 resolution) overlaid with a 0.04 mm thick UV filter sheet (allowing less than $10\%$ light transmission below 390 nanometers) for visual stimulus presentation, a UMA-8 multichannel USB microphone array V2.0 (miniDSP) for audio monitoring and recording, acoustic absorption panels for soundproofing and temperature and humidity sensors. A Raspberry Pi 4 model B was used to conduct data collection and stimulus presentation in each chamber. We monitored vocalizations continuously during daylight hours in all recording chambers and saved recorded audio on a common data server for offline parsing and analysis. A 15-minute long pre-recorded video of female conspecifics was presented once a day to stimulate singing behavior~\cite{galoch_behavioural_2007, carouso-peck_female_2019, ikebuchi_male_1999}. Audio recordings were automatically parsed using custom, Python-based signal processing algorithms to track the number of songs present in each bird's daily vocalizations. Subjects that yielded the highest number of song renditions per day were selected for neural implantation~\cite{yamahachi_undirected_2020}. The birds were not used in any other experiments.

\subsection*{\hypertarget{methods:surgical}{Surgical procedure}} 
Into selected birds, we implanted a single Neuropixels 1.0 probe~\cite{jun_fully_2017} targeting the telencephalic regions HVC and RA. All surgical procedures were performed under general anesthesia (gaseous mixture of isoflurane/oxygen $1\%$-$2\%$) and aseptic conditions. Birds were administered antibacterial enrofloxacin (Enrosite; $20 \unit{mg}/\unit{kg}$ I.M.) on the 2 days prior to implant surgery and a preoperative injection of butorphanol tartrate (Turbogesic; 2 $\unit{mg}/\unit{kg}$ I.M.) the day of the procedure. Carprofen (Rimadyl; 4 $\unit{mg}/\unit{kg}$ I.M.) was administered as necessary for postoperative analgesia on the days following the implant. Water-diluted antibiotics (Baytril) were provided for at least seven days post-surgery. Birds were head-fixed in a digital stereotaxic frame (Leica Biosystems Inc.). Head feathers were plucked, and betadine was applied over the scalp. A straight incision from the top of the beak to the back of the head was performed to expose the cranium, and the upper layer of the skull over the Y-sinus was removed. A reference wire made with a 0.5 \si{\milli\meter} segment of platinum-iridium wire ($0.002"$ thick) soldered to a 1 \si{\milli\meter} segment of silver wire ($0.005"$ thick) was placed between the dura and the skull through a small craniotomy centered at 3 \si{\milli\meter} contralateral to the hemisphere were the Neuropixels probe was implanted. The head of the bird was angled at $\ang{130}$, measured as the angle formed between the horizontal axis and the line forming from the bird's beak to the bird's ear when head-fixed in the stereotaxic apparatus. This step, informed by the zebra finch brain atlas, was critical to align the HVC and RA nuclei in the dorsal-ventral axis. The cortex of the brain was exposed by a craniotomy (centered at 2500 \si{\micro\meter} lateral and 300 \si{\micro\meter} anterior to Y-sinus) in the hemisphere of interest. A durotomy was performed. A 0.5 \si{\mega\ohm}, monopolar tungsten microelectrode (Science Products Co.) was inserted at the durotomy site and neural activity was monitored under anesthesia using an Intan RHD USB Interface Board, reaching a ventral depth of 3000 \si{\micro\meter}. The identification of characteristic HVC and RA neural patterns validated the accuracy of the implant's targeted placement. A Neuropixels probe, secured by an in-house designed holding shaft, was implanted at the target location to a depth of 3 \si{\milli\meter} to ensure coverage of both HVC and RA regions. The shank of the Neuropixels probe was stained with Dil fluorescent dye ($DilC_{18}(3)$) prior to implantation for post-mortem histology trace tracking. The holding shafts of the Neuropixels probe and Neuropixels Headstage, designed for stabilizing the probe and headstage during neural recordings, were affixed to the skull using adhesive cement (C$\&$B Metabond \textregistered). The durotomy was subsequently covered with dural substitute silicon gel (3-4680) (Dow Corning Corp.), and the head skin was sutured back in place. The overall added weight was 2.2 \si{\gram} $\pm$ 0.2 \si{\gram}. To confirm the implantation site of the Neuropixels probe in HVC and$\slash$or RA and validate the recorded neural activity, a post-mortem histological assessment was conducted (\hyperref[S2]{Sup. Fig. 2}). All birds presented in this study were implanted in the right brain hemisphere.

\subsection*{\hypertarget{methods:experiment}{Experiment Design and Neural Recordings}}
A commercial single-housing acoustically-isolated chamber (Eckel Noise Control Technologies Co.) was used for awake, chronic recordings. Simultaneous neural and behavioral (song) data recordings started twelve to twenty-four hours post-surgery. The additional weight of the implant (Neuropixels probe and headstage) to the bird's head was supported by a weight reliever mechanism consisting of a counterweight mass of ~1.5 $\si{\gram}$ routed through a pulley using a thin nylon wire (0.008-$\si{inch}$ diameter fishing line). Each recording session comprised 120 \si{\minute} to 240 \si{\minute} of continuously recorded data as the birds moved around and sang freely. Song was recorded either in the presence of a female conspecific, indicating putative directed song for bird A and bird B, or in the absence of such, indicating putative undirected song for bird C and bird D. To capture the spiking activity of HVC and RA neurons, the voltage signals recorded by 384 Neuropixels channels were amplified, band-pass filtered (300 \si{\hertz}-10000 \si{\hertz}), multiplexed and digitized at 30 \si{\kilo\hertz} on the base (Neuropixels headstage) and directly transferred to a PXIe acquisition module (Imec R\&D Co.) contained in a PXIe-1071 chassis (National Instruments Corp.) that provided a high-bandwidth backplane. Audio signals were recorded via a TC20 omnidirectional microphone (Earthworks Audio Co.), sampled at 40 \si{\kilo\hertz} via a BNC-2090A analog breakout accessory (National Instruments Corp.) and high-pass filtered (250 \si{\hertz}) for subsequent extraction of spectral features. Neural and behavioral data were synchronized via a common 1 \si{\hertz} clock signal generated by the PXIe Acquisition module. 

\subsection*{\hypertarget{methods:histology}{Histological assessment}}
The brain was carefully isolated after the end of an experiment and immediately immersed in a 10\% Neutral Buffered Formalin (NBF) solution. To ensure proper histology, the brain was transferred to a 30\% w/v sucrose solution (30 \si{\g} of sucrose in 100 \si{\milli\liter} of distilled water) for cryoprotection 72 hours before the procedure. Prior to histology, the brain was rinsed with PBS. The specimen was embedded in Tissue-Tek O.C.T. compound onto the surface of a microtome and prepared for sagittal sectioning. While the microtome was maintained at ($-27^\circ C$), dry-iced was used to help freeze the tissue quickly and efficiently for subsequent frozen sectioning. 40-70 \si{\micro\meter} slices were cut and mounted on individual glass slides. Tissue sections were examined under a light microscope and images were captured using a digital camera.

\subsection*{\hypertarget{methods:neural_processing}{Neural Signal Processing}}
\begin{figure}[h]
\label{S1}
\centering  
\includegraphics[width=1.0\textwidth]{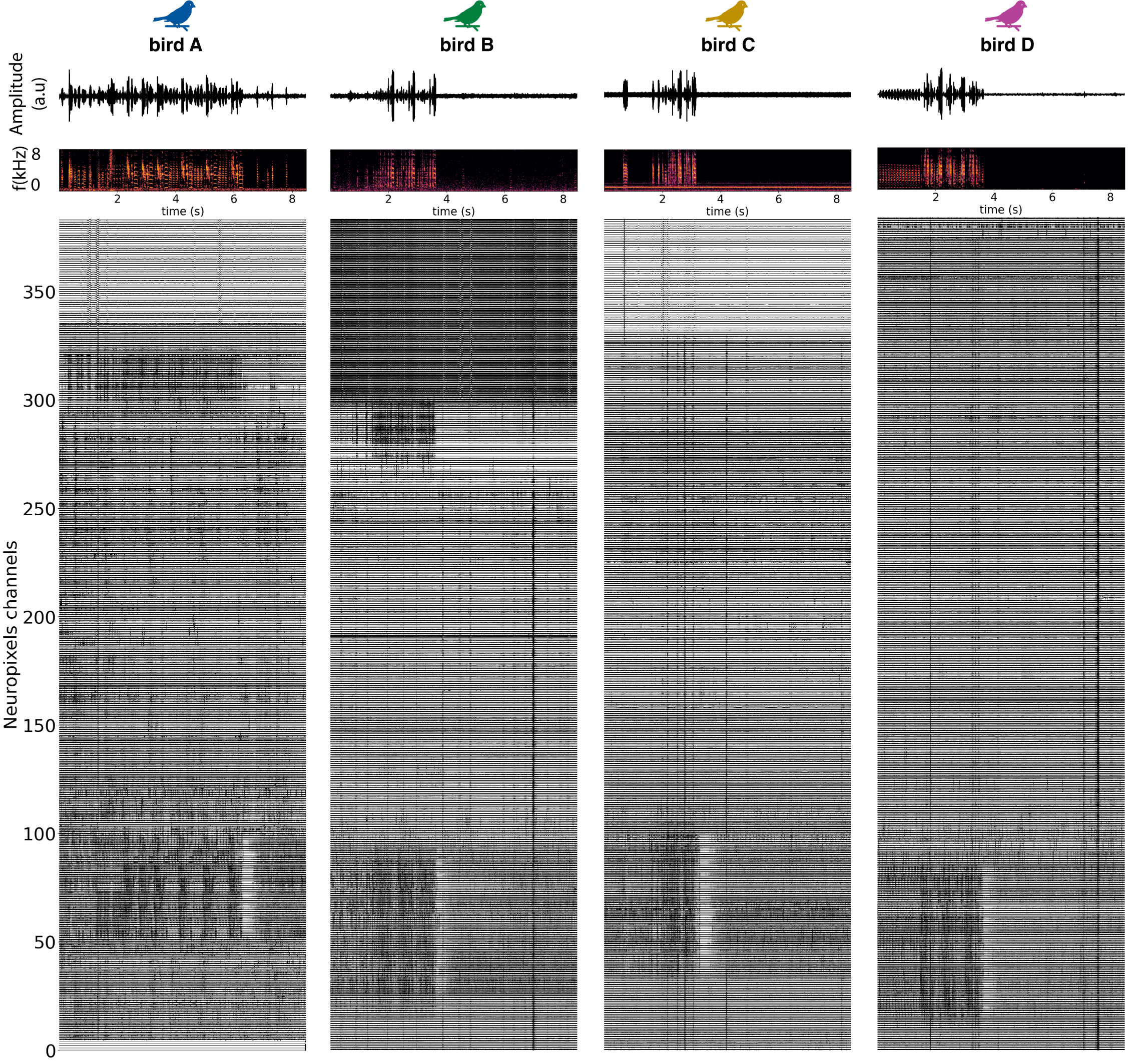}
\end{figure}

\begin{figure}[h]
\centering  
\includegraphics[width=1.0\textwidth]{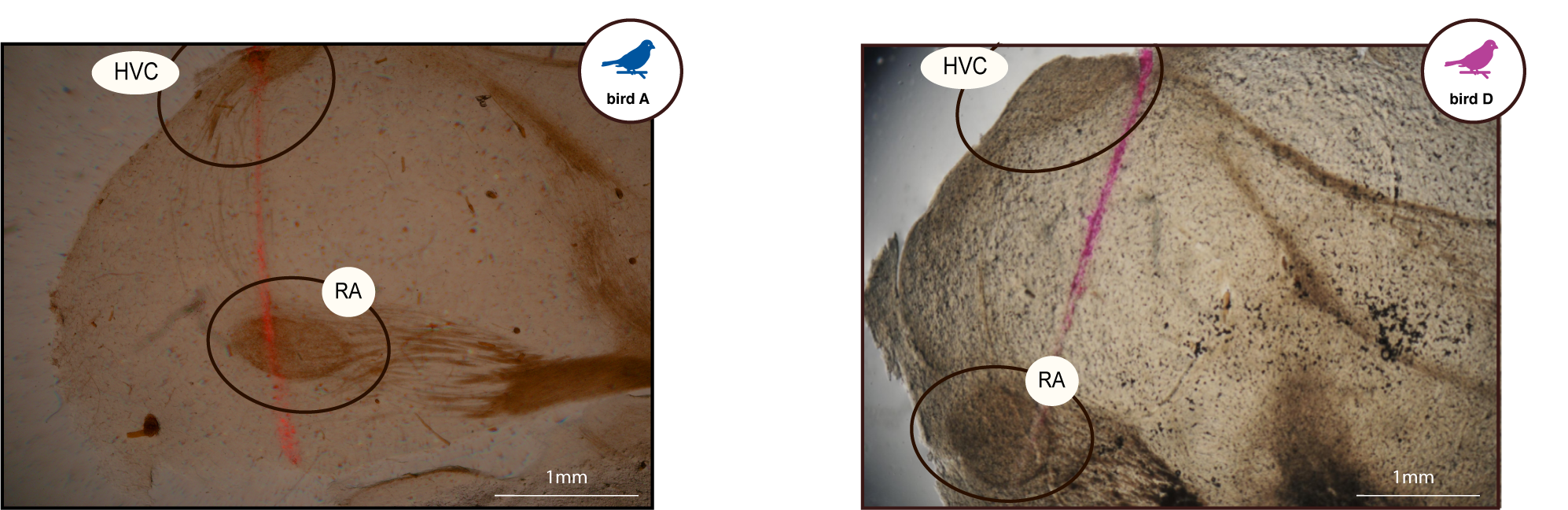}
\label{S2}
\end{figure}

\begin{figure}[h]
\label{S3}
\centering  
\includegraphics[width=1.0\textwidth]{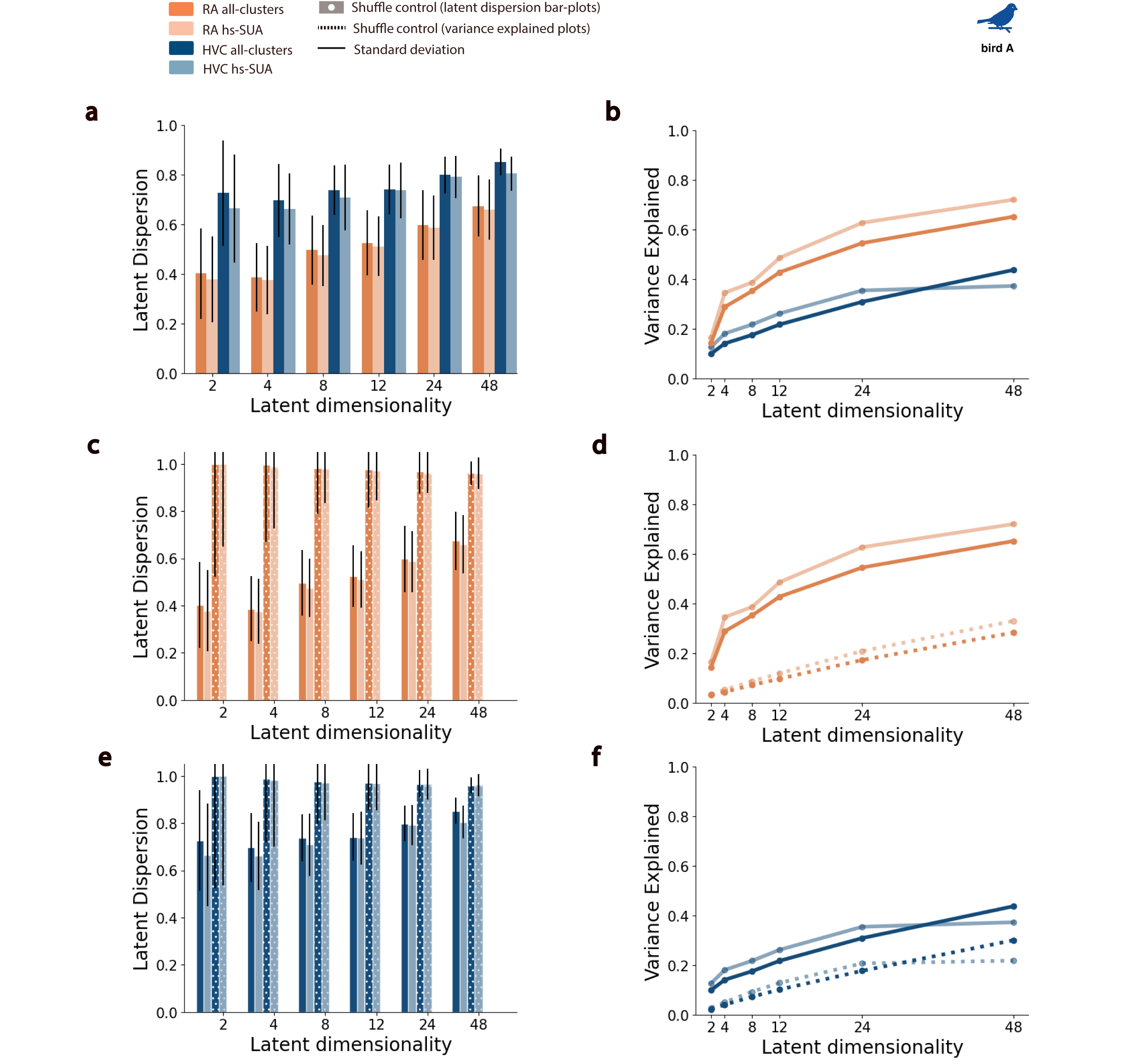}
\end{figure}

\begin{figure}[h]
\centering  
\includegraphics[width=1.0\textwidth]{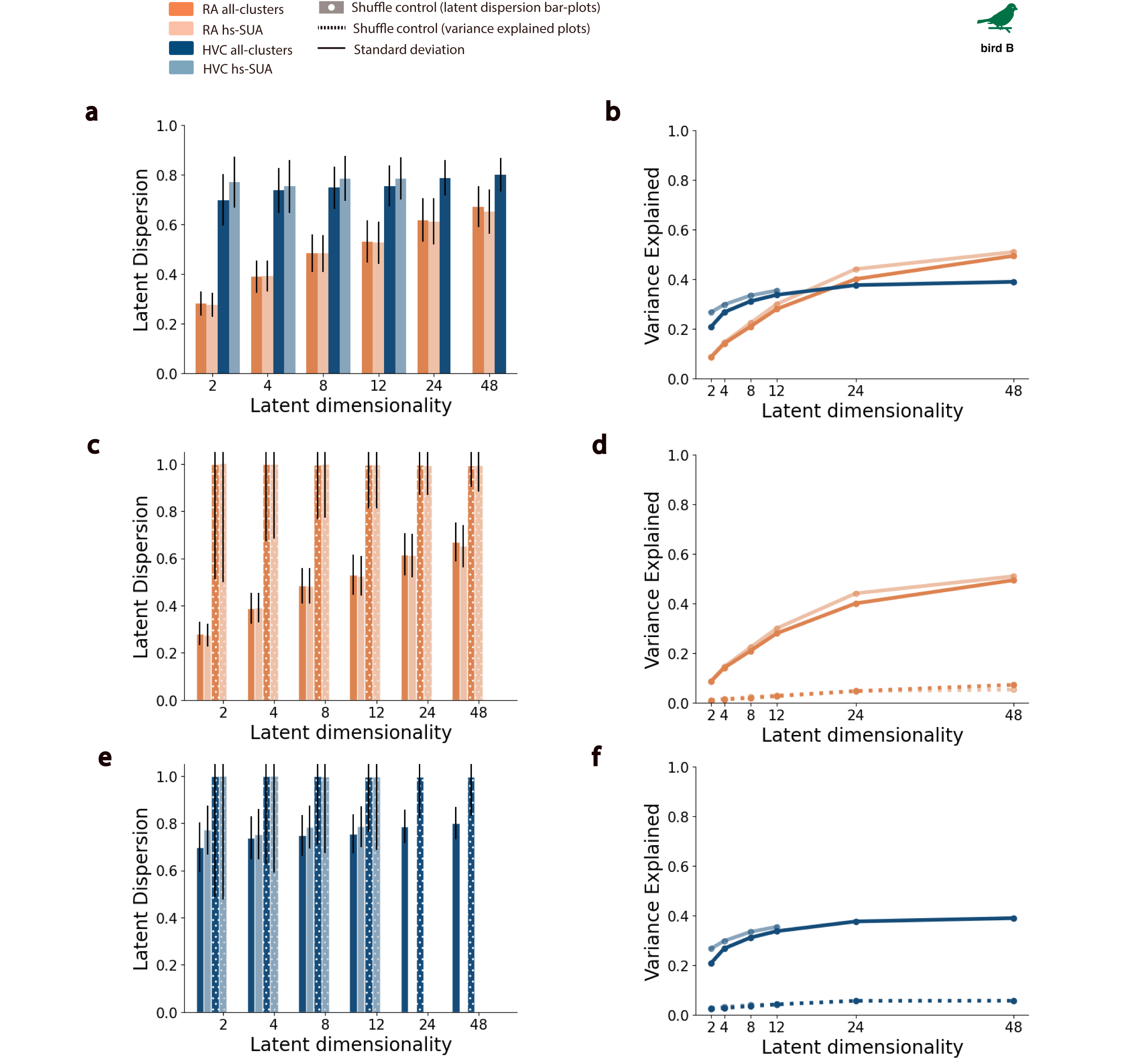}
\label{S4}
\end{figure}

\begin{figure}[h]
\centering  
\includegraphics[width=1.0\textwidth]{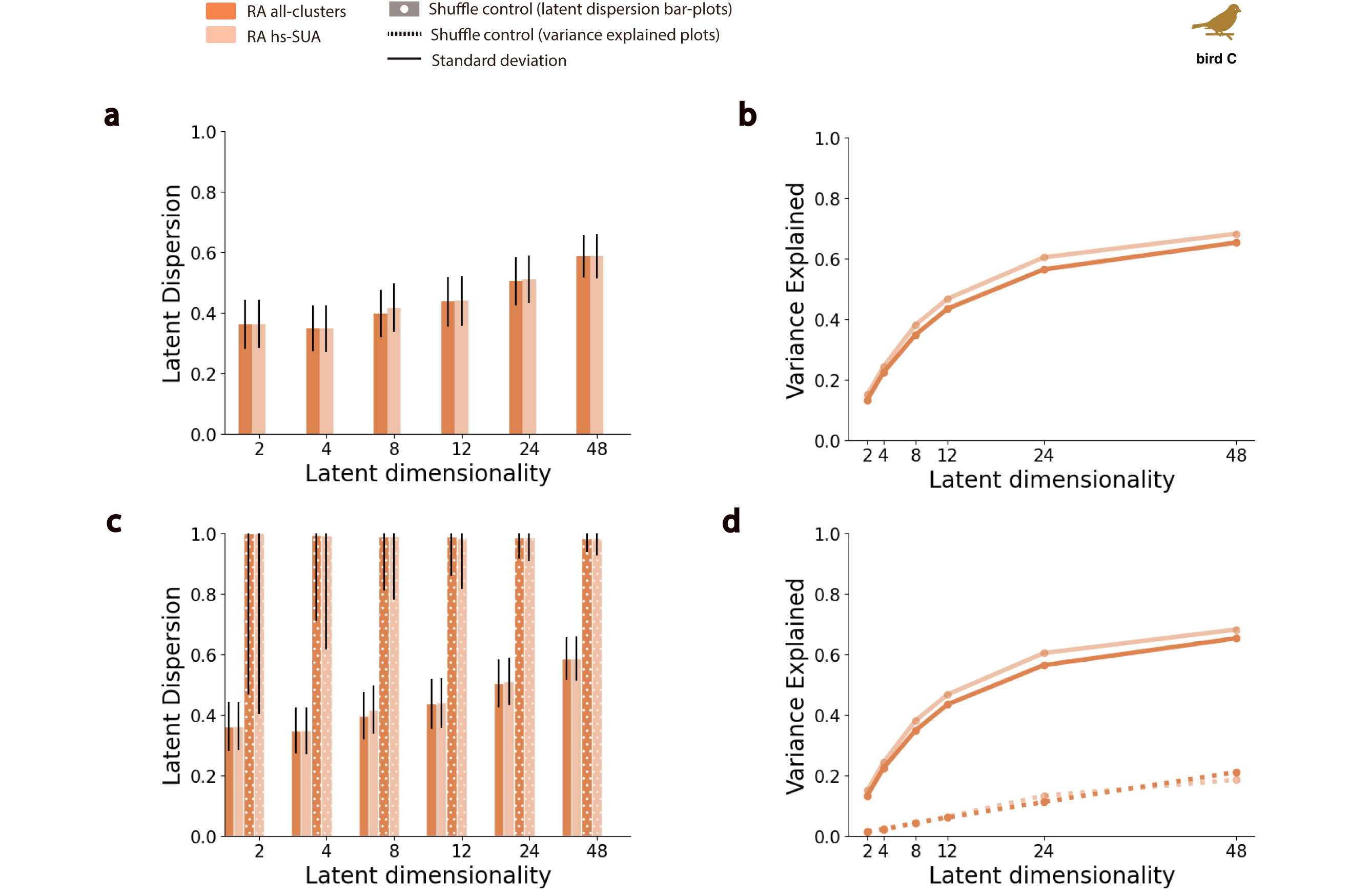}
\label{S5}
\end{figure}

\begin{figure}[h]
\centering  
\includegraphics[width=1.0\textwidth]{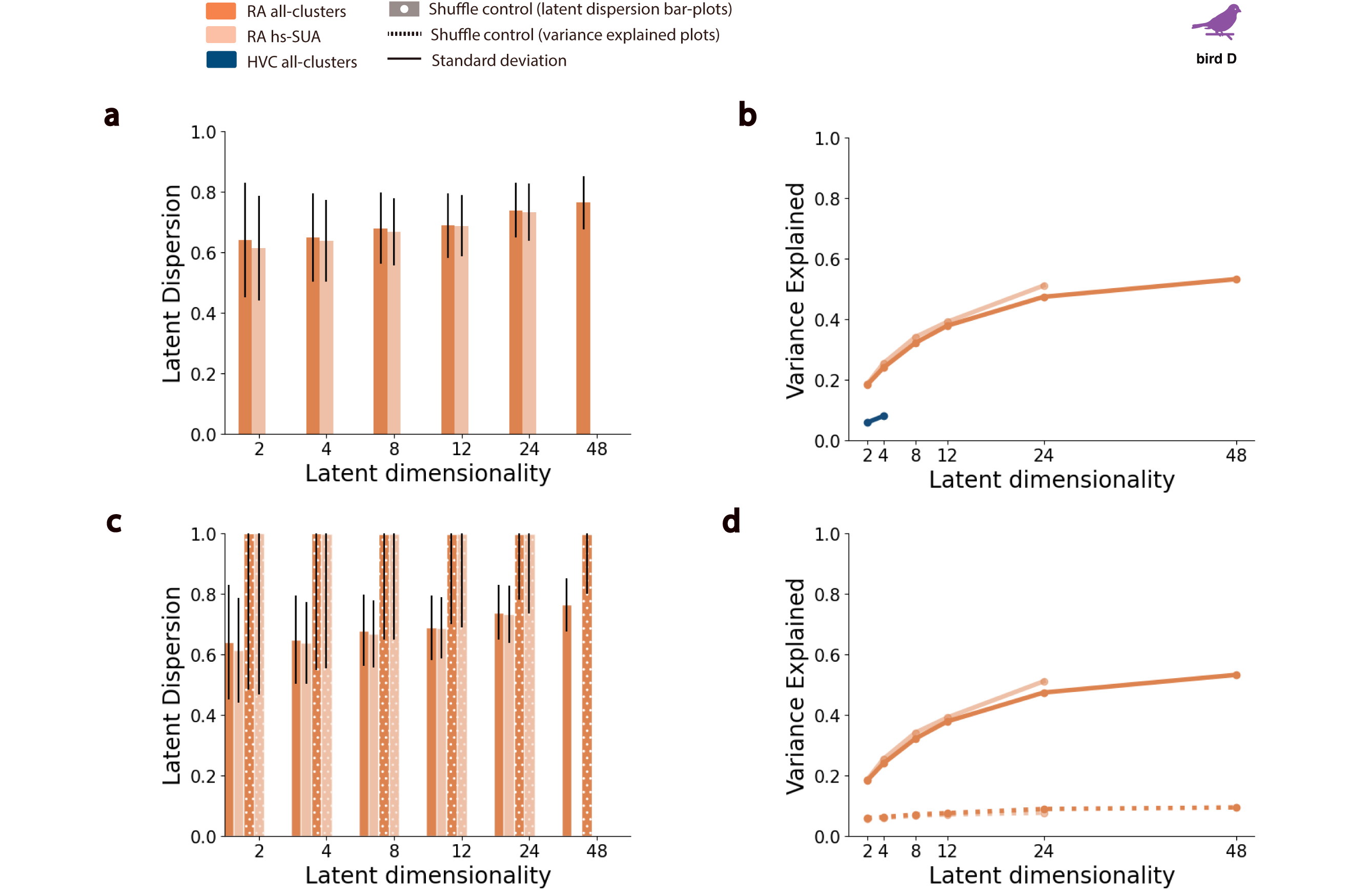}
\label{S6}
\end{figure}

\begin{figure}[h]
\centering  
\includegraphics[width=1.0\textwidth]{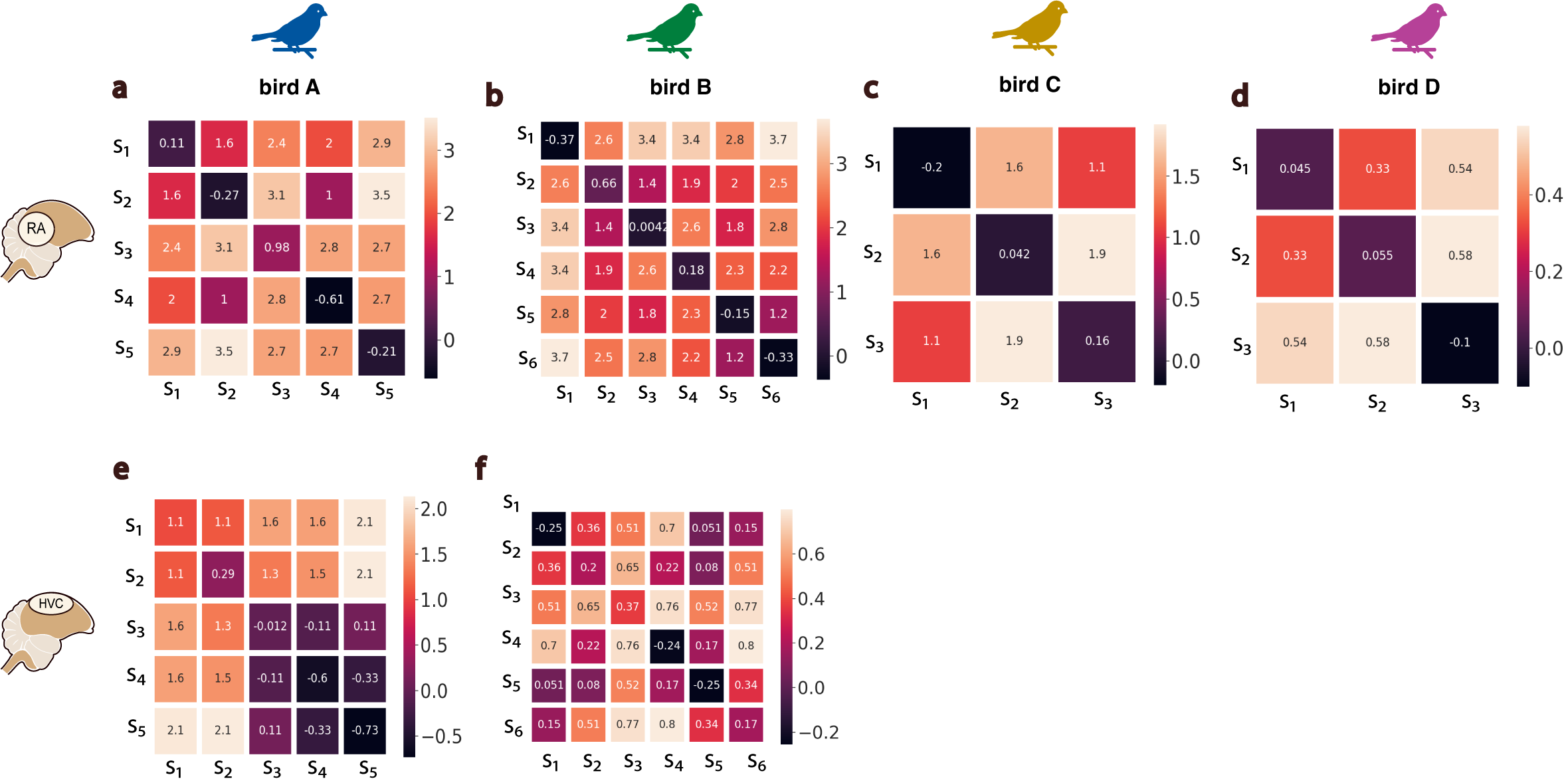}
\label{S7}
\end{figure}

Spike detection and clustering were performed using Kilosort3 software~\cite{pachitariu_solving_2023} via SpikeInterface~\cite{buccino_spikeinterface_2020} and manual post-processing (merging and$\slash$or splitting of clusters) was performed using Phy 2.0. Neural clusters that were identified were categorized into high-quality single-unit activity (SUA) clusters and multi-unit activity (MUA) clusters. This classification was determined by assessing the quality of isolation, which was based on several factors including the extracted waveform, the distribution of spike amplitudes, the consistency of spike detection throughout the recording, autocorrelogram analysis, and the fraction of refractory period (2 \si{ms}) violations in the inter-spike interval (ISI) distribution (with a requirement of <3\% for SUA clusters). Clusters that were visibly noisy (as per the action potential waveform or the distribution of spike times throughout the recording session) were excluded from further analysis. Clusters were designated as putative HVC or RA clusters if they exhibited spiking modulation during song production and if their primary recording channel was located within the anticipated group of recording channels for either target brain area, as expected from the observed raw traces during the recordings (\hyperref[S1]{Sup. Fig. 1}). Histological assessment of the probe location confirmed these expectations. Premotor HVC SUA clusters were inspected and labeled as putative sparse-firing HVC projection neurons ($HVC_{pn}$) or tonically firing HVC interneurons ($HVC_{int}$). A SUA cluster in HVC was labeled as a putative ($HVC_{pn}$) if its mean, spontaneous firing rate was below 5 \si{\hertz} during periods of song and presented temporally local bursts of activity at 100 \si{\hertz} or higher during motifs~\cite{kozhevnikov_singing-related_2007, rauske_state_2003}. Spike counts for each sorted cluster were then collapsed into 1 \si{ms} time bins (30 samples at 30 \si{ksps}). Although birds tended to perch still during periods of vocal activity, body movement and head shaking could cause occasional motion artifacts that were reflected in the neural data streams. These artifacts might have produced the erroneous, simultaneous detection of spikes across electrode channels. To reduce the effect of such artifacts, we identified near-simultaneous spiking activity by summing all spike counts across channels at a 1 \si{ms} time-resolution. Spiking events where the across-channels, combined spike count was $>5$ S.D. above the mean of the distribution were removed from all clusters.

\subsection*{Behavioral parsing and annotation}
Periods of song were automatically identified within each session using custom motif-based template-matching algorithms and curated manually as described previously~\cite{arneodo_neurally_2021}. Syllables, motifs and other vocalizations during periods of song were annotated by experienced researchers using Praat software~\cite{boersma_speak_2001}. Manual annotations were used to identify onsets and offsets of song bouts and individual song syllables and extract time-aligned spike times accordingly.

\subsection*{\hypertarget{methods:dynamics}{Neural population latent dynamics}}
Data analysis was performed utilizing Python 3 custom software along with open-source libraries (\hyperlink{methods:code}{Code Availability}). 

We used state-space modeling techniques to find latent neural manifolds that captured the covariance structure of the activity of the neural populations of interest. Specifically, we applied Gaussian-Process Factor Analysis (GPFA) to spike trains of $HVC$ and $RA$ populations of neurons recorded during birdsong. GPFA provides a probabilistic framework aimed at extracting smooth neural trajectories from individual experimental trials through a combination of smoothing and dimensionality-reduction operations~\cite{yu_gaussian-process_2009}. A linear-Gaussian relationship is assumed between observations (spike trains: $y_{:,t} \in \mathbb{R}^{qx1}$) and latent neural states ($x_{:,t} \in \mathbb{R}^{px1}$) as follows: $y_{:,t} | x_{:,t} = \mathcal{N}(Cx_{:,t} + d, \space R)$, where $q$ denotes the number of neurons recorded and $p$ denotes the assumed dimensionality of the underlying latent space, or neural manifold. $C \in \mathbb{R}^{qxp}$, $d \in \mathbb{R}^{qx1}$ and $R \in \mathbb{R}^{qxq}$ are model parameters fitted using the expectation-maximization (EM) algorithm to maximize the likelihood of the observed data. The covariance matrix, $R$, is constrained to be diagonal, where the diagonal elements correspond to the independent noise variances of each neural cluster. Neural states across adjacent time points are related through Gaussian processes, which aim to enforce smoothness throughout the neural trajectories. An independent Gaussian process was defined for each latent dimension, where the chosen form of the covariance matrix relating adjacent neural states was the squared exponential covariance function~\cite{yu_gaussian-process_2009}. Song bouts were identified within each recording session. Trials were generated by splitting song bouts into non-overlapping segments of vocal activity around the stereotyped portions of song motifs. Spike trains of different combinations of SUA and MUA clusters were used to train independent GPFA models for each session. To assess the dimensionality of the neural manifold associated with the vocal task of interest in each neural population, trajectories were computed to exist in $d$-dimensional latent spaces after binning the spike trains into 15 \si{\milli\second} bins $d \in [2, 4, 8, 16, 32, 48]$~\cite{sadtler_neural_2014, gallego_long-term_2020}. To map the resulting latent states ($x_{:,t}$) onto an orthonormal space, an orthonormalization procedure was carried out, involving the application of the singular value decomposition of the learned linear mapping matrix ($C$). This step is necessary to allow for an intuitive visualization of the learned latent trajectories and to identify a set of orthonormal basis vectors (spanning the same space as the original columns of $C$) ordered by the amount of data covariance explained~\cite{yu_gaussian-process_2009}. Through this procedure, orthonormalized neural states ($\tilde{x}_{:,t}$) are calculated, corresponding to the latent neural modes that explain the most variance of the original spike trains. 

GPFA assumes a Gaussian distribution on the square-root counts of the input spike trains. This assumption poses a mathematical limitation particularly on the spike trains of $HVC_{pn}$, which are characterized by a highly bursty, non-gaussian spiking distribution. To investigate the effectiveness of our methods to identify relevant neural manifolds among populations of neurons with diverse spiking characteristics, we selected neural subpopulations within $HVC$ and $RA$ according to the following criteria. On the one hand, we identified a subpopulation of neurons characterized by their clear isolation (high-quality SUA clusters) and an average firing rate exceeding $10\unit{Hz}$ during periods of song. In the context of $HVC$, this subset of neurons is expected to omit the proportion of $HVC_{pn}$ and/or $HVC_{int}$ that exhibit markedly sparse, bursty firing rates which may not adhere to the assumptions of our models. We term this subpopulation \textit{hs-SUA} (i.e. high spiking single-unit activity). On the other hand, we trained separate state-space models on all identified clusters within our recorded population, including both tonically and sparsely firing SUA and MUA clusters. We refer to this subpopulation as \textit{all-clusters} (\hyperref[fig:fig4]{Figure 4}). To evaluate and compare the significance of the estimated neural manifolds in relation to song production and to assess trial-to-trial variability within the latent space, we employed the following metrics.

\textbf{Latent dispersion}. We introduced a metric named \textit{latent dispersion} ($\bar{\sigma}$) to quantify the spread of trial-specific neural trajectories around the average trajectory during consistent overt behavior. For any given trial, latent dispersion ($\sigma_n$) is calculated as the mean deviation of each neural state ($\vec{x}_{n,t}  \in \mathbb{R}^{px1}$) from the average trajectory $\vec{\mu}_t = \frac{1}{N} \sum_{n=1}^{N} \vec{x}_{n,t}$ over time steps ($T$). To enable the comparison of latent dispersion measurements across different neural manifolds, such as those of HVC and RA, latent dispersion values are normalized against the overall variance within the latent space, i.e. the mean distance from neural states to the center of the spanned neural manifold $\vec{\mu} = \frac{1}{T} \sum_{t=1}^{T} \vec{\mu}_t$. As a result, the normalized latent dispersion for a trial ($\sigma_n$) within a d-dimensional latent manifold is expressed as:

\begin{equation}
\sigma_n = \sum_{t=1}^T \frac{\| \vec{x}_{n,t} - \vec{\mu}_t \|}{\| \vec{\mu}_t - \vec{\mu} \|}
\end{equation}

We report the average and standard deviation latent dispersion across trials ($N$) for each identified neural manifold, which provides intuition about how stable the neural trajectories are for repetitive behavior in each latent space.

\textbf{Variance Explained}. The amount of shared neural variability captured by the latent neural trajectories is calculated as the ratio between the shared variance, defined by the loading matrix ($C$), and the sum of the shared variance and the independent (private) variances of each neuron ($R$):
\begin{align}
&\text{Total Variance} = tr(C * C' + R)\\
&\text{Variance Explained (shared)} =  tr(C * C') / \text{Total Variance}\\
&\text{Private Variance} = tr(R) / \text{Total Variance}
\end{align}

\subsection*{\hypertarget{methods:controls}{Sub-population Control Analyses}}
To determine whether factors such as population size or neuron-spiking properties (for instance, sparsity of neural firing) influenced the observed outcomes, we conducted a series of control analyses. To account for the discrepancy in the population size sampled from RA in comparison to HVC, control experiments adjusting for neuron-count were conducted. These involved fitting state-space models to randomly selected subsets of RA clusters, ensuring the neuron counts matched those of the respective HVC populations recorded from each bird (\hyperref[fig:fig4]{Figure 4}). The average spikerate of the recorded HVC populations exceeded that of the corresponding subsampled RA populations, thus eliminating the potential influence of total spike-count on our results. To further investigate whether our findings were affected by the presence of particularly sparse neural clusters or the quality of neuronal isolation, we performed experiments comparing neural manifolds derived exclusively from high-spiking, well-isolated single-unit activity (hs-SUA) clusters to those inferred from all identified single-unit (SUA) and multi-unit activity (MUA) clusters (all-clusters). For each control experiment conducted, we report the variance explained as a function of latent dimensionality and latent dispersion metrics within the inferred manifolds, providing a direct comparison to the findings from the original experiments.

\textbf{Significance Analysis}. To assess the statistical significance of our findings, we conducted a bootstrap analysis comparing 12-dimensional GPFA models fitted to different neural populations, across birds. This involved 10 bootstrap samples (fitted models) per subgroup, each consisting of 90\% randomly chosen putative HVC and RA neurons from distinct neural populations (all-clusters and hs-SUA). The selection of the manifold's dimensionality was strategically made to accommodate the varying number of isolated neurons from each population and bird, thereby maximizing the inclusion of neural populations across birds in our analysis. Subsequently, we performed a Tukey's HSD test to compare the mean variance explained and the mean latent dispersion across neural manifolds fitted to HVC and RA neural subpopulations, following the identification of a significant F-statistic in an Analysis of Variance (ANOVA) across subpopulations.

\subsection*{\hypertarget{methods:distance}{Latent manifold distance analysis}}
To explore the continuity of the HVC and RA neural manifolds and how they relate to vocal outputs, we conducted the following within and across-manifold distance analyses. Initially, we identified the latent neural states corresponding to the onset of each motif syllable along the neural trajectories. Subsequently, we calculated two distinct distance distributions within each neural manifold: one for pairwise distances between states related to the onset of identical syllables across motifs, named the \textit{same-syllable distance} distribution ($SSD_M$), and another for pairwise distances between states related to the onset of dissimilar syllables across motifs, named the \textit{dissimilar-syllable distance} distribution ($DSD_M$), where $M$ denotes the specific manifold of interest (HVC or RA).

\textbf{Within manifold distance analysis}. To determine whether neural states associated with the production of the different vocalizations are separable within the inferred neural manifolds, we treated the \textit{same-syllable distance} distribution ($SSD_M$) as reference. We then normalized both the ($SSD_M$) and ($DSD_M$) distributions with respect to this reference as follows:

\begin{equation}
Z_M^* = \frac{*_M - mean(SSD_M)}{std(SSD_M)}, \quad\quad * = \{SSD, DSD\}.
\end{equation}

Within each manifold, we used a Welch’s t-test on the unequal-sized distributions to assess the statistical significance of the difference between the \textit{same-syllable} ($Z_M^{SSD}$) and the \textit{dissimilar-syllable} ($Z_M^{DSD}$) distance distributions.

\textbf{Between-manifold distance analysis}. To quantify the difference in the effects observed between the HVC and RA neural manifolds, we conducted a Wilcoxon signed-rank test comparing the means of their normalized \textit{dissimilar-syllable distance} distributions ($Z_{HVC}^{DSD}$ versus $Z_{RA}^{DSD}$), with each sample in these distributions being directly matched on a one-to-one basis by design (\hyperref[fig:fig5]{Figure 5}).

\subsection*{\hypertarget{methods:branching}{Vocalization branching analysis}}
To investigate whether the latent trajectories described by the HVC and RA population activity encoded non-deterministic syllable transitions in the observed vocal output, we identified “branch points” in each birdsong where the syllable sequencing was probabilistic. The variability in syllable sequencing typical of Zebra finch song is illustrated in the behavioral state diagrams shown in (\hyperref[fig:fig6]{Figure 6}). A representative example is bird A, presenting 7 different song syllables (syllables 1-6 and i) and 3 “branch points” (syllables i, 5 and 6) where sequencing is probabilistic rather than deterministic. For example, syllable 5 may be followed by syllables 1, 6 or silence. We annotated motif renditions based on the type of transition that occurred after the highlighted “branch point” and asked whether we could predict the occurring vocal transition type from the sequence of HVC and RA latent states. We trained a Fisher's linear discriminant (LDA) to estimate the posterior probability of the correct vocal rendition  type from neural states at each time point along the latent trajectory. The class probability priors were assumed to be uniform ($1/C$), where $C$ is the number of potential transitions after the “branch point”. We conducted a bootstrap analysis (1000 runs) in which, in each run, we evaluated the model's performance using a leave-one-out cross-validation scheme with a random $80\%-20\%$ train-test partition of the data. We ensured that at least one example of each class was present in the randomly selected training partition. We report the continuous mean correct-class posterior probability with $95\%$ confidence intervals (\hyperref[fig:fig6]{Figure 6}).

\subsection*{\hypertarget{methods:embeddings}{EnSongdec: Song embeddings extraction through EnCodec}}
The Pytorch-based machine learning pipeline developed for this project incorporated proprietary software to design and train custom neural networks, leveraged open-source libraries to access pre-trained models available from Hugging Face and utilized Weights \& Biases for experiment tracking (\hyperlink{methods:code}{Code Availability}). 

In our vocal decoder designs, we utilized the EnCodec audio model to extract embedded representations of birdsong recordings~\cite{defossez2022highfi}. EnCodec, a state-of-the-art audio codec pre-trained on music, employs an encoder-decoder architecture with a quantized latent space to achieve high-fidelity audio compression with minimal reconstruction loss. The model comprises three primary components:

(i) The encoder network (E), which features a combination of 1D convolutional layers and LSTM units, takes an audio sample $x$ and transforms it into a 128-dimensional floating-point audio latent representation $z$.

(ii) A quantization layer (Q) then converts $z$ into a compressed latent representation $z_q$ utilizing residual vector quantization (RVQ).

(iii) The decoder network (G) mirrors the encoder's structure but uses transposed convolutions to reconstruct the original time-domain signal $\hat{x}$ from $z_q$.

To prioritize reconstruction quality in our experiments, we opted for minimal audio compression settings, specifically 24kbps at a 48kHz input, corresponding to using 16 residual quantization steps with each step containing 1024 entries (10 bits per codebook). We used the encoder network (E) to extract meaningful embedding representations of birdsong in our datasets.

\subsection*{\hypertarget{methods:network}{EnSongdec: Neural-to-song embeddings network}}

We trained custom feed-forward neural networks to translate neural signals into continuous latent representations of birdsong. Initially, song extracts were mapped to their corresponding song embedded representations ($z$) using the EnCodec’s Encoder network. Subsequently, a feed-forward network was optimized to decode neural inputs into these song-embedding representations. Models were independently trained to reconstruct vocal outputs from both spike-trains and neural latents derived from HVC or RA neural activities. Decoding from spike-sorted neural recordings involved smoothing spike-trains using a 1d-Gaussian kernel ($\sigma=30$) and downsampling to match the sampling rate of the target audio embeddings (150 samples per second). For neural latents, latent trajectories were inferred from spike-sorted neural recordings using GPFA at a $5$ $\unit{ms}$ temporal resolution in a 12-dimensional neural manifold and similarly downsampled. The resulting feed-forward neural network architecture featured an input layer of size $i = N \times history\_bins$ (where $N$ denotes the number of isolated clusters in models decoding from spike-sorted neural data, and $N = 12$ for models decoding from neural trajectories). History bins was set to three bins of neural history (15 \unit{ms}-20 \unit{ms}) preceding the output bin. This was followed by two 64-unit hidden layers and a 128-unit output layer that produced a 1x128 array, representing the predicted audio embedding corresponding to every bin of neural data. ELU activation functions were employed in the input and hidden layers. Models were optimized to reconstruct audio embeddings using mean square error (MSE) as the reconstruction loss.

\subsection*{\hypertarget{methods:training}{EnSongdec: Training Procedure}}
Data was organized into non-overlapping song motifs (audio recordings) and their corresponding neural traces. For each bird and recording session, 80\% of all identified motifs were randomly selected for training, with the remainder used for testing. A 10\% subset of the training data was reserved for validation to aid hyperparameter optimization. AdamW optimizer and an MSE loss function were employed. Dropout of 30\% was implemented in the input and hidden layers during training. Temporal jitter (up to $10\unit{ms}$) and Gaussian noise ($\mu=0, \sigma=0.25$) were added probabilistically ($p=0.5$) as data augmentation strategies to mitigate overfitting. Multiple experiments were conducted to refine the feed-forward network architecture that best converged across conditions.

\subsection*{\hypertarget{methods:stability}{EnSongdec: Latent Stability Analysis}}
To simulate a scenario involving a complete turnover of the recorded neurons within a population, we conducted experiments to assess the stability of decoders trained on the spike-rates of individual neurons and those trained on inferred neural latents. We split the HVC and RA populations into two equal halves ($Z_1$ and $Z_2$), and computed the corresponding spike-rates and neural trajectories, using GPFA, for each split. We then trained decoders on one half of the neural data ($Z_1$) following our predefined \hyperlink{methods:training}{training scheme}. To test these decoders in simulated scenarios where acute neural shift occurs, we evaluated their performance on the 20\% test sets using the second partition ($Z_2$) of the neural data as neural input instead. Given that the neural spaces of $Z_1$ and $Z_2$ (either high-dimensional or GPFA-derived latent manifolds) are likely to be misaligned, we employed Canonical Correlation Analysis (CCA)  to identify a canonical space where the linear combinations of $Z_1$ and $Z_2$ show maximum pairwise correlation. This well-established linear algebra-based method has been effectively used in previous studies of motor cortex data~\cite{canonical_2003, gallego_cortical_2018, sussillo_neural_2015, farshchian_adversarial_nodate}. After aligning $Z_2$ to the $Z_1$ neural space through this canonical space, we assessed the performance of our EnSongdec models on this adjusted $Z_2$ data (\hyperref[fig:fig7]{Figure 7g,h}).

\subsection*{\hypertarget{methods:synthesis}{EnSongdec: Continuous synthesis of naturalistic vocalizations}}
\begin{figure}[h]
\centering  
\includegraphics[width=1.0\textwidth]{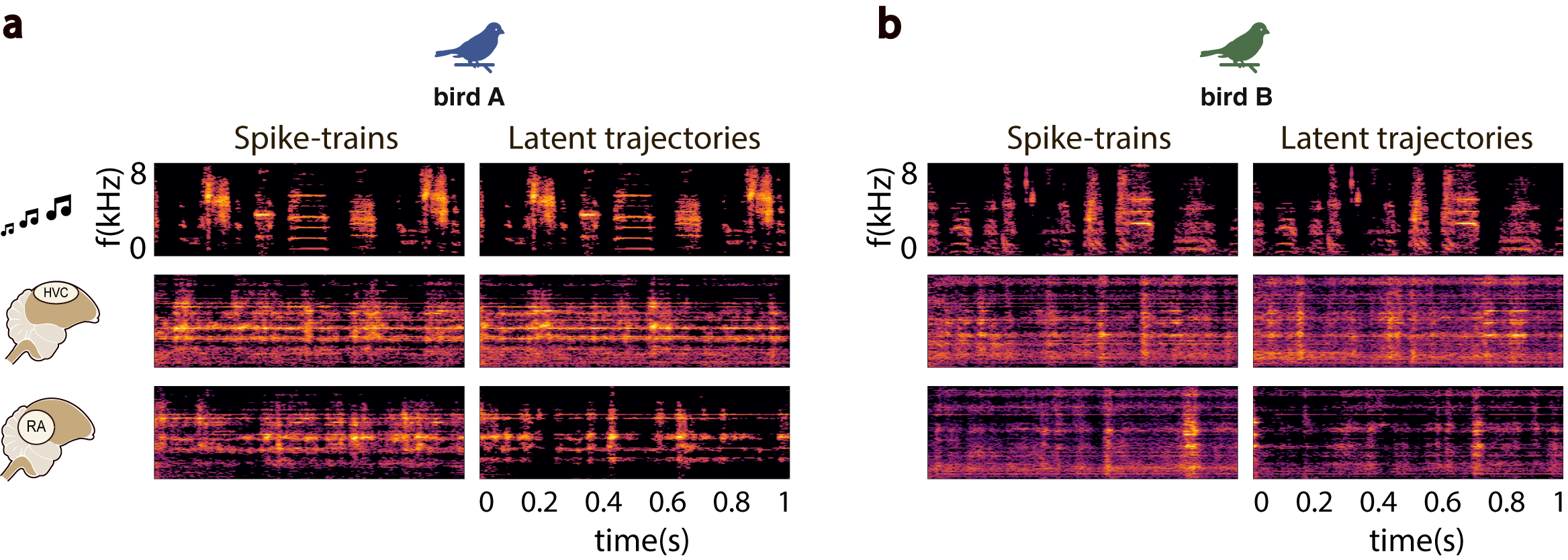}
\label{S8}
\end{figure}

EnSongdec comprises a neural-to-song embedding network that translated neural activity (either as spike rates or neural latent trajectories) into latent representations of birdsong. These representations were then processed by the pre-trained EnCodec’s Quantizer-Decoder network to reconstruct the time-domain song signal. The performance of the models was evaluated as the pixel-wise mean-squared error between the normalized spectrograms of the reconstructed and original song signals in the testing set (\hyperref[fig:fig7]{Figure 7}). Controls involving the random shuffling of neural inputs to the network along the temporal dimension were conducted to confirm that the high-fidelity reconstruction of the song was not artifactual (\hyperref[S8]{Sup. Fig. 8}).

\clearpage
\section*{Supporting information}

\paragraph*{Figure S1.}
{\bf  Examples of Neuropixels recordings time-locked to the production of song bouts in zebra finches.} Each panel presents the audio waveform and corresponding spectrogram of a sample song bout for four different zebra finch exemplars. Below each, the corresponding neural traces recorded synchronously using a single Neuropixels probe are shown. RA population activity is identifiable in all four recordings, marked by a notable suppression of neural activity at the termination of the song bout. Birds A and B include HVC population activity. Bird C recordings solely capture RA activity. Bird D recordings contain only a few Neuropixels channels capturing HVC activity, insufficient for population-based analyses.

\clearpage
\paragraph*{Figure S2.}
{\bf Histological images of Neuropixels implants in zebra finches.} Bird A’s histological image confirmed a successful implant across HVC and RA target regions. In contrast, the histological image for Bird D showed that while RA was successfully targeted, HVC was only tangentially targeted, resulting in only a few Neuropixels channels recording HVC activity.

\clearpage
\paragraph*{Figure S3.}
{\bf Bird A - Dimensionality-Dependent Analysis of Latent Neural Manifolds.} \textbf{a.} Mean latent dispersion in GPFA-inferred neural manifolds as a function of latent dimensionality across \textit{hs-SUA} and \textit{all-clusters} HVC and RA neural populations. \textbf{b.} Neural variance explained by GPFA-inferred neural manifolds as a function of latent dimensionality across \textit{hs-SUA} and \textit{all-clusters} HVC and RA neural populations. \textbf{c.} Mean latent dispersion in GPFA-inferred neural manifolds as a function of latent dimensionality in RA neural populations and corresponding neural shuffle controls. \textbf{d.} Neural variance explained by GPFA-inferred neural manifolds as a function of latent dimensionality in RA neural populations and corresponding neural shuffle controls. \textbf{e.} Mean latent dispersion in GPFA-inferred neural manifolds as a function of latent dimensionality in HVC neural populations and corresponding neural shuffle controls. \textbf{f.} Neural variance explained by GPFA-inferred neural manifolds as a function of latent dimensionality in HVC neural populations and corresponding neural shuffle controls.

\clearpage
\paragraph*{Figure S4.}
{\bf Bird B - Dimensionality-Dependent Analysis of Latent Neural Manifolds.} \textbf{a.} Mean latent dispersion in GPFA-inferred neural manifolds as a function of latent dimensionality across \textit{hs-SUA} and \textit{all-clusters} HVC and RA neural populations. \textbf{b.} Neural variance explained by GPFA-inferred neural manifolds as a function of latent dimensionality across \textit{hs-SUA} and \textit{all-clusters} HVC and RA neural populations. \textbf{c.} Mean latent dispersion in GPFA-inferred neural manifolds as a function of latent dimensionality in RA neural populations and corresponding neural shuffle controls. \textbf{d.} Neural variance explained by GPFA-inferred neural manifolds as a function of latent dimensionality in RA neural populations and corresponding neural shuffle controls. \textbf{e.} Mean latent dispersion in GPFA-inferred neural manifolds as a function of latent dimensionality in HVC neural populations and corresponding neural shuffle controls. \textbf{f.} Neural variance explained by GPFA-inferred neural manifolds as a function of latent dimensionality in HVC neural populations and corresponding neural shuffle controls.

\clearpage
\paragraph*{Figure S5.}
{\bf Bird C - Dimensionality-Dependent Analysis of Latent Neural Manifolds.} \textbf{a.} Mean latent dispersion in GPFA-inferred neural manifolds as a function of latent dimensionality across \textit{hs-SUA} and \textit{all-clusters} RA neural populations. \textbf{b.} Neural variance explained by GPFA-inferred neural manifolds as a function of latent dimensionality across \textit{hs-SUA} and \textit{all-clusters} RA neural populations. \textbf{c.} Mean latent dispersion in GPFA-inferred neural manifolds as a function of latent dimensionality in RA neural populations and corresponding neural shuffle controls. \textbf{d.} Neural variance explained by GPFA-inferred neural manifolds as a function of latent dimensionality in RA neural populations and corresponding neural shuffle controls. 

\clearpage
\paragraph*{Figure S6.}
{\bf Bird D - Dimensionality-Dependent Analysis of Latent Neural Manifolds.} \textbf{a.} Mean latent dispersion in GPFA-inferred neural manifolds as a function of latent dimensionality across \textit{hs-SUA} and \textit{all-clusters} RA neural populations. \textbf{b.} Neural variance explained by GPFA-inferred neural manifolds as a function of latent dimensionality across \textit{hs-SUA} and \textit{all-clusters} RA neural populations. \textbf{c.} Mean latent dispersion in GPFA-inferred neural manifolds as a function of latent dimensionality in RA neural populations and corresponding neural shuffle controls. \textbf{d.} Neural variance explained by GPFA-inferred neural manifolds as a function of latent dimensionality in RA neural populations and corresponding neural shuffle controls. 

\clearpage
\paragraph*{Figure S7.}
{\bf Syllable-based Latent Neural State Distance Confusion Matrices.} 
Confusion matrices showing normalized, mean Euclidean distances in latent manifold across neural states corresponding to the production of different syllables in \textbf{a.} bird A - Ra \textbf{b.} bird B - RA \textbf{c.} bird C - RA \textbf{d.} bird D - RA \textbf{e.} bird A - HVC \textbf{f.} bird B - HVC.

\clearpage
\paragraph*{Figure S8.}
{\bf EnSongdec Controls - Synthesis of Audio Spectrograms from Shuffled Neural Activity.} Synthesised spectrograms using HVC and RA shuffled spike-trains and GPFA-inferred latent states along the temporal dimension as inputs to EnSongdec in \textbf{a.} bird A and \textbf{b.} bird B.

\clearpage


\section*{Ethics Statement}
Procedures and methods comply with all relevant ethical regulations for animal testing and research and were carried out in accordance with the guidelines of the Institutional Animal Care and Use Committee at the University of California, San Diego (protocol S15027).

\section*{Acknowledgments}
We thank members of the Gilja and Gentner Labs for the continuous feedback to refine the manuscript. We especially acknowledge Dr. Aashish Patel, Dr. Tim Sainburg and Sophia Huang for critical discussions and suggestions for improving the manuscript. We thank Michael Turvey for guidance on histology procedures. We thank Dr. Caterina Morganti for insightful input on the artwork. We thank Dr. Aashish Patel for his invaluable help to maintain the computing infrastructure used during this project. 

\section*{Funding}
This work was supported by the NIH National Institute on Deafness and Other Communication Disorders (grants R01DC018446 and R01DC008358), the NSF Emerging Frontiers in Research and Innovation (EFRI) - Brain-Inspired Dynamics for Engineering Energy-Efficient Circuits and Artificial Intelligence (BRAID) (grant 2223822) and the Kavli Institute for Brain and Mind (IRG no. 2021-1759). P.T.M. acknowledges support from "La Caixa" Foundation and the IIE Fulbright Fellowship. D.E.B acknowledges support from the University of California—Historically Black Colleges and Universities Initiative and the National Science Foundation Graduate Research Fellowship DGE-1650112. X.A.P. acknowledges support from the GEM Fellowship and the Summer Training Academy for Research Success (STARS) Fellowship. Any opinion, findings and conclusions or recommendations expressed in this material are those of the authors(s) and do not necessarily reflect the views of the funders. The funders had no role in study design, data collection and analysis, decision to publish, or preparation of the manuscript.

\section*{Author contributions}
P.T.M., E.M.A, T.Q.G. and V.G. conceptualized the project. P.T.M., E.M.A. and L.O. performed the experiments and collected the data. P.T.M., E.M.A., X.A.P. and A.K. built the hardware and software in the homemade recording chambers. E.M.A built the experimental rig and wrote the software for the song pipeline. P.T.M. and E.M.A. wrote the software for the neural data pipeline. P.T.M., E.M.A., D.E.B., L.L.S. and A.A. pre-processed the data and annotated the behavior. P.T.M handled the data analysis, designed the brain-to-song decoder architecture and conducted the song synthesis experiments. P.T.M. wrote the manuscript. All authors discussed and edited the manuscript. T.Q.G. and V.G. jointly supervised the work.

\section*{Competing Interests}
V.G. holds shares in Neuralink, Corp., and Paradromics, Inc. and currently consults for Paradromics, Inc. These organizations had no role in study design, data collection and analysis, decision to publish, or preparation of the manuscript.

\clearpage
\bibliography{arxiv_manifolds}

\end{document}